\documentclass{jfm}
\usepackage{graphicx}
\usepackage{epstopdf, epsfig}

\shorttitle{Magneto-Rayleigh-Taylor instability in an elastic medium}
\shortauthor{S. A. Piriz, A. R. Piriz and N. A. Tahir}

\title{Magneto-Rayleigh-Taylor instability in an elastic finite-width medium overlying an ideal fluid}

\author{S. A. Piriz\aff{1}, A. R. Piriz\aff{1}
  \corresp{\email{roberto.piriz@uclm.es}},
 \and N. A. Tahir\aff{2}}

\affiliation{\aff{1}Instituto de Investigaciones Energ\'{e}ticas (INEI), E.T.S.I.I., and CYTEMA, Universidad de Castilla-La Mancha, 13071 Ciudad Real, Spain 
\aff{2}GSI Helmholtzzentrum f\"{u}r Schwerionenforschung Darmstadt, Planckstrasse 1, 64291
Darmstadt, Germany}

\begin{document}

\maketitle

\begin{abstract}
We present the linear theory of two-dimensional incompressible magneto-Rayleigh-Taylor instability in a system composed of a linear elastic (Hookean) layer above a lighter semi-infinite ideal fluid with magnetic fields present, above and below the layer. As expected, magnetic field effects and elasticity effects together enhance the stability of thick layers. However, the situation becomes more complicated for relatively thin slabs, and a number of new and unexpected phenomena are observed. In particular, when the magnetic field beneath the layer dominates, its effects compete with effects due to elasticity, and counteract the elasticity stabilising effects. As a consequence, the layer can become more unstable than when only one of these stabilising mechanism is acting. This somewhat unexpected result is explained by the different physical mechanisms for which elasticity and magnetic fields stabilise the system. Implications for experiments on magnetically driven accelerated plates and implosions are discussed. Moreover, the relevance for triggering of crust-quakes in strongly magnetised neutron stars is also pointed out.
\end{abstract}


\section{Introduction}
Rayleigh-Taylor (RT) instability is a well known phenomenon in hydrodynamics that occurs whenever a denser medium lays on the top of a lighter one in a uniform gravitational field $\boldsymbol{g}$ or, equivalently, when the denser medium is pushed and accelerated by the lighter one with an acceleration $-\boldsymbol{g}$ (Rayleigh 1883, Taylor 1950). This instability has been widely studied for the case of semi-infinite media (Chandrasekhar 1961), but much less attention has been paid to the cases involving finite-thickness media, especially when these media are not in contact with rigid walls. In fact, the presence of rigid surfaces reduces the number of boundary conditions and simplifies the mathematical treatment considerably. This has usually been the case when considering the RT instability in more complex situations as, for instance, when viscous fluids (Mikaelian 1982), or elastic media (Mora et al. 2014, Ricobelli \& Ciarletta 2017) are involved. 

However, in most of the experiments on high energy density physics (Barnes et al. 1974, Kalantar et al. 2000, Park et al. 2010), applications to inertial confinement fusion (Davies et al. 2017, Seyler et al. 2018), as well as in the contexts of astrophysics (Blaes et al. 1990, 1992, Mock et al. 1998), and Earth and planetary science (Burov \& Molnar 2008, Tahir et al. 2006, 2017, 2018), the heavier medium is in contact with lighter fluids or it has free surfaces. This is also the scenario in some recent laboratory experiments (Adkins et al. 2017) involving viscous fluids, a situation that has also been studied theoretically by Piriz, Piriz \& Tahir (2018) and previously, for some particular limits, by Lister \& Kerr (1989), and Wilcock \& Whitehead (1991). For the case of finite-width elastic-media with no presence of rigid boundaries RT instability has been studied by Bakhrakh et al. (1997), Plohr \& Sharp (1998), and by Piriz, Piriz \& Tahir (2017a, 2017b).

When magnetic fields are present the instability is known as the magneto-Rayleigh-Taylor (MRT) instability. To our knowledge, the case concerning finite-thickness media has only been studied when the involved media are ideal fluids or plasmas (Harris 1962, Lau et al. 2011). In some research, a viscous fluid has also been considered, but it was assumed to be limited by rigid walls (Awasthi 2014). On the other hand, for the MRT instability involving an elastic medium, it seems to have been studied only for the simplest configuration of two semi-infinite media (Sun \& Piriz 2014). 

However, the more interesting situation in which the heavy medium is a slab with elastic properties, is of great relevance to many experiments on high-energy-density physics involving magnetically accelerated flyer plates that still retain its mechanical properties when it is impacted on a target sample (Lemke et al. 2011, Martin et al. 2012). In addition, this problem is of interest in the recently proposed approach to inertial confinement fusion known as magnetic inertial fusion, in which a magnetic field is used to mitigate the thermal conduction losses from the compressed fusion fuel, so that the ignition requirements are relaxed (Davies et al. 2017, Seyler 2018). The presence of such an interior magnetic field will also play a role in the implosion stability, especially when the initial field becomes compressed during the implosion and its intensity is considerably increased. 

Besides, MRT in elastic media may also be of importance in the crust-quakes taking place in the strongly magnetised neutron stars known as magnetars, which are considered to be at the origin of the emissions of soft $\gamma$-rays and of the $X$-rays pulsars (Cheng et al. 1995, Kaminker et al. 2009). In fact, it has been shown that pycnonuclear and electron capture reactions forced by the mass accretion from the interstellar medium can lead to the development of a density inversion in the crust of the neutron star (Blaes et al. 1990,1992, Mock \& Joss 1998, Bildsten \& Cumming 1998). However, in order that such a density inversion can drive the RT instability in the neutron star crust, its magnitude must exceed some minimum value imposed by the crust elasticity (Blaes et al. 1990, Piriz, Piriz \& Tahir 2017b). Since the maximum density inversion, as determined by the dynamics of the crust, has been shown to be unable to reach such a minimum (Mock \& Joss 1998), the crust will remain stable unless the stabilising effect of elasticity can be somehow reduced. 

It is not at all intuitive that the presence of magnetic fields may alter this scenario by eliminating the instability threshold imposed by the elasticity of the crust. Especially if we take into account that the addition of magnetic fields and elasticity effects in semi-infinite media leads to the enhancement of the system stability (Sun \& Piriz 2014). However, we show in this work that when a sufficiently thin elastic slab is considered, a competition phenomenon takes place between magnetic fields and elasticity for which the effectivity of the elasticity is progressively diminished as the magnetic field intensity increases until the instability threshold completely disappears. This competition phenomenon may become an issue for the magnetic inertial fusion aiming to use solid slabs in combination with magnetic fields to mitigate the effects of the MRT instability during the acceleration process.
\begin{figure}
\centering
\begin{center}
\vspace*{-1.2 mm}
\includegraphics[width= 0.75 \textwidth, clip]{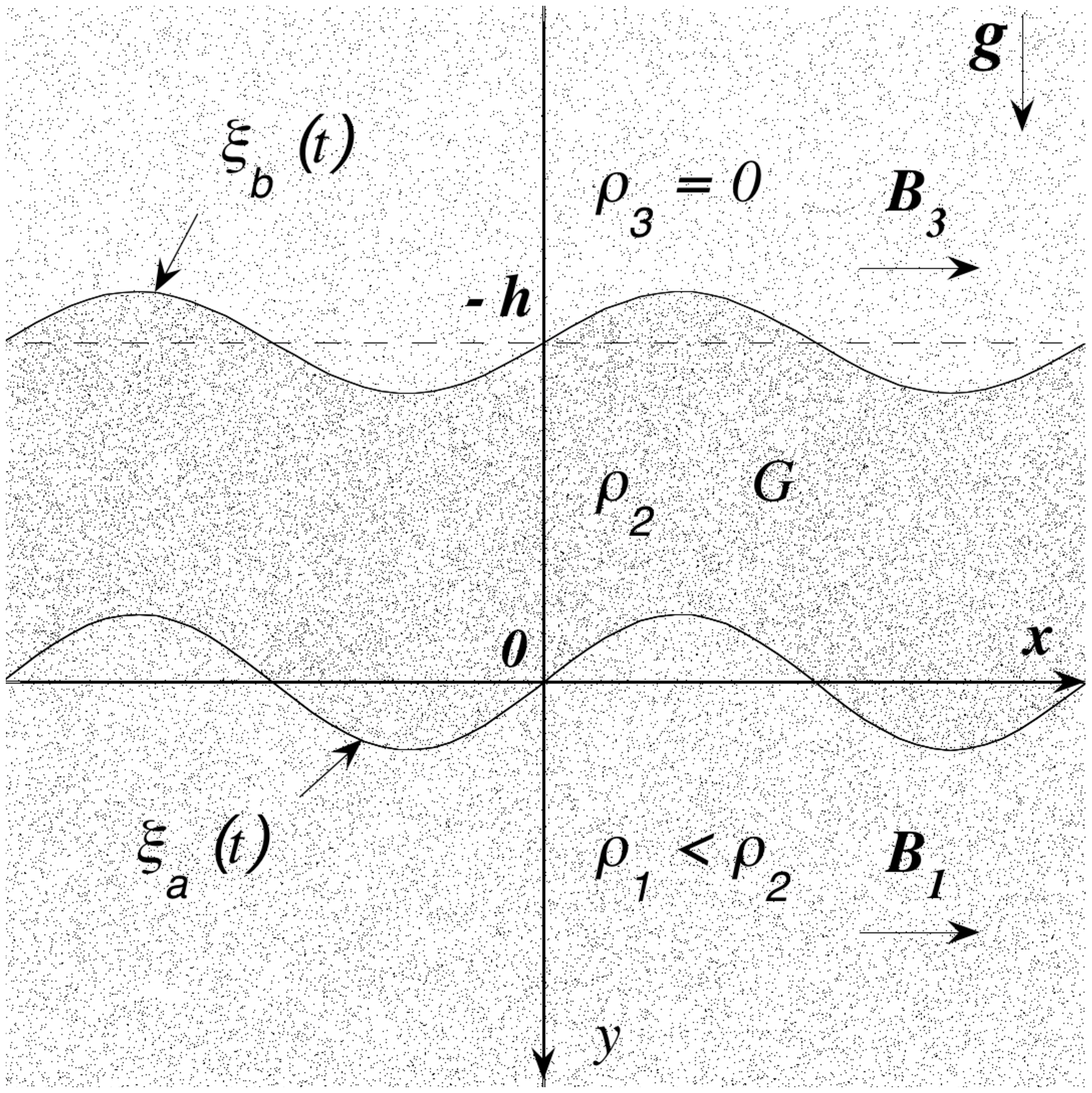}
\end{center}
\vspace*{-50 mm}
\caption{Schematic of the two corrugated interfaces system formed by an incompressible elastic slab with average thickness $h$ laying above a lighter semi-infinite incompressible ideal fluid with horizontal magnetic fields at both sides of the slab.}
\label{Figure-1}
\end{figure}

\section{Linear MRT instability}
\subsection{Fundamental equations}
We consider the situation of the two-dimensional system schematically described in figure 1, in which a Hookean medium of thickness $h$, density $\rho_2$, and shear modulus $G$, occupies the magnetic-field-free region $-h\leq y \leq 0$. The slab overlays an ideal fluid of density $\rho_1 < \rho_2$ occupying the region $y \geq 0$, which is filled with a uniform magnetic field $\boldsymbol{B_1}=B_1 \boldsymbol{e_x}$ ($\boldsymbol{e_x}$ is the unitary vector in the direction of the $x$-axis). In the region over the slab, $y\leq -h$, we assume a medium with density $\rho_3=0$ (physically, it would be a tenuous ideal medium such that $\rho_3 \ll \rho_2$, $\rho_1$) that is also filled with a uniform magnetic field $\boldsymbol{B_3}=B_3 \boldsymbol{e_x}$. The whole system is under the action of the uniform gravity acceleration field $\boldsymbol{g}=g \boldsymbol{e_y}= -\boldsymbol{\nabla} \varphi$ ($\boldsymbol{e_y}$ is the unitary vector in the vertical direction, and $\varphi$ is the gravitational potential). In each region we consider that the medium is incompressible, perfectly conducting, and that there are no free charges. The media are also considered to be immiscible.

We start the analysis of the MRT instability with the equations for mass and momentum conservation in the following general form:
\begin{equation}
\frac{d \rho_n}{dt} + \rho_n \frac{\partial v_{ni}}{\partial x_i}=0
\ ,\label{1}
\end{equation}
\begin{equation}
\rho_n \frac{dv_{ni}}{dt} = - \frac{\partial p_n}{\partial x_i} +\rho_n g \delta_{iy} + \frac{\partial \sigma_{ik}^{(n)}}{\partial x_k}  + \epsilon_{ijk} J_{nj} B_{nk} 
\ ,\label{2}
\end{equation}
where $n=1,2,3$ refer, respectively, to the bottom, middle and top regions, and we have used index notation for cartesian vectors and tensors so that $i =1, 2,3$ indicate,  the space coordinates $x$, $y$, or $z$. Furthermore $\epsilon_{ijk}$ is the Levi-Civita pseudo-tensor [$\epsilon_{ijk} J_{nj} B_{nk} 
= (\boldsymbol{J_n \times B_n})_i$], $\boldsymbol{J_n }$ is the current density, $\boldsymbol{B_n }$ is the magnetic field as defined above in the region $n$, $v_{ni}$, $\rho_n$, and $p_n$ are, respectively, the $i^{th}$ velocity component, density, and pressure; and $\sigma_{ik}^{(n)}$ is the deviatoric part of the stress tensor $\Sigma_{ik}^{(n)} = -p_n \delta_{ik} +\sigma_{ik}^{(n)}$ of the medium $n$ ($\delta_{ik}$ is the Kronecker $\delta$). We will use interchangeably vector and index notation as appropriate for the calculations presentation. Besides, $d \Theta/dt$ represents the material derivative of any magnitude $\Theta$:
\begin{equation}
\frac{d \Theta}{dt}= \frac{\partial \Theta}{\partial t} + (\boldsymbol{v \cdot \nabla}) \Theta
\ .\label{4}
\end{equation}

The previous equations have to be complemented with the Maxwell equations. Namely, the current density $\boldsymbol{J_n}$ is related to the magnetic field $\boldsymbol{B_n}$ by means of the Ampere's law. In a medium with no free-charges, it reads:
\begin{equation}
\boldsymbol{\nabla \times B_n}=\mu_n \boldsymbol{J_n}
\ ,\label{5}
\end{equation}
where $\mu_n$ is the magnetic permeability on the $n$-medium. Then, the Lorentz's force $\boldsymbol{F_{Ln}}=\boldsymbol{J_n \times B_n}$ is:
\begin{equation}
\boldsymbol{F_{L}^{(n)}} =\frac{(\boldsymbol{\nabla \times B_n}) \times \boldsymbol{B_n}}{\mu_n}
\ ,\label{6}
\end{equation}
which can be written in terms of the magnetic stress tensor $M_{ik}$ by using the Gauss's law for the magnetism $\boldsymbol{\nabla \cdot B_n}=0$:
\refstepcounter{equation}
$$
F_{Li}^{(n)}=\frac{\partial M_{ik}^{(n)}}{\partial x_k}, \quad
M_{ik}^{(n)}= - \frac{B_{nj} B_{nj}}{2 \mu_n} \delta_{ik}+ \frac{B_{ni} B_{nk}}{\mu_n} \ .
  \eqno{(\theequation{\mathit{a},\mathit{b}})} \label{7}
$$
On the other hand, the magnetic field $\boldsymbol{B_n}$ is related to the material velocity $\boldsymbol{v_n}$ through the Faraday's law:
\begin{equation}
\frac{\partial \boldsymbol{B_n}}{\partial t}= -\boldsymbol{\nabla \times E_n}= \boldsymbol{\nabla \times(v_n \times B_n)}
\ ,\label{8}
\end{equation}
where we have taken into account the Ohm's law $\boldsymbol{J_n}= \eta_n (\boldsymbol{E_n} + \boldsymbol{v_n \times B_n})$ ($\eta_n$ is the electrical conductivity) for media with very high conductivity ($\eta_n \rightarrow \infty$) or with $\boldsymbol{J_n}=0$, so that we have $\boldsymbol{E_n}=- \boldsymbol{v_n \times B_n}$. The previous equation can also be written in the following alternative form:
\begin{equation}
\frac{d B_{ni}}{d t}= \frac{\partial v_{ni}}{\partial x_j} B_{nj}
\ ,\label{8a}
\end{equation}
which shows that the magnetic field is "frozen" in the fluid and it moves as a material substance (Chandrasekhar 1961). 

From (\ref{7}$b$), the momentum conservation equation (\ref{2}) can be re-written in a more symmetric form as follows:
\begin{equation}
\rho_n \frac{dv_{ni}}{dt} = - \frac{\partial p_n}{\partial x_i} +\rho_n g \delta_{iy} + \frac{\partial (\sigma_{ik}^{(n)}+ M_{ik}^{(n)}  )}{\partial x_k}  
\ .\label{7a}
\end{equation}
In addition, for a Hookean medium $\sigma_{ik}^{(n)}$ is given by the following expression:
\begin{equation}
\frac{\partial \sigma_{ik}^{(n)}}{\partial t}= G_n \left( \frac{\partial v_{ni}}{\partial x_k} +  \frac{\partial v_{nk}}{\partial x_i}\right)
\ .\label{3}
\end{equation}

 Then, we can solve the linear stability problem by considering small amplitude perturbations about the solution for a flat slab. Since in equilibrium the deviatoric part of the stress tensor vanish, and we are assuming an incompressible medium, we obtain the following hydrostatic solution:
 \begin{equation}
p(y) =  \left \{ \begin{array}{ll}
                                     0 \  \  \   \  \  \  \  \   \  \  \  \   \  \  \  \  \   \  \  \  \   \  \  \   \  y \leq -h
				 \\
				\\ 
                                  \frac{B_{3}^{2}}{2\mu_3} + \rho_2g(y+h)  \  \  \  -h \leq y \leq 0
				 \\
				\\ 
				 p_1  + \rho_1gy, \  \   \  \   \  \  \   \  \  \     \  \  \     y \geq 0 
	\end{array} \right.  \ \ ,  \label{4a}
\end{equation}
where $p_1=  \rho_2gh+ \frac{B_{3}^{2}}{2\mu_3} - \frac{B_{1}^{2}}{2\mu_1} $. 

In order to close the problem we need to impose the boundary conditions corresponding to our physical situation in which the top surface of the linear elastic slab is a free surface in contact with an empty region filled with the horizontal uniform magnetic field $\boldsymbol{B_3}$, while the bottom surface is in contact with an ideal fluid which is embedded in the uniform magnetic field  $\boldsymbol{B_1}$. Integrations of (\ref{1}) and (\ref{7a}) along the vertical coordinate $y$, across the bottom and the top interfaces respectively, yield the following jump conditions representing the continuity of the normal velocities, and of the normal and tangential stresses at such interfaces:
\begin{equation}
\| - p_n \delta_{ik} +\sigma_{ik}^{(n)} n_k+ M_{ik}^{(n)} n_k \| =0 
\ ,\label{4b}
\end{equation}
\begin{equation}
\| -v_{ny} \| =0 
\ ,\label{4c}
\end{equation}
(with $i= y$ or $x$) where $\|Q\|=Q(y_0+0^+)-Q(y_0-0^+)$, with $y_0=0$ or $y_0=-h$, and $0^+ \rightarrow 0$. 

 \subsection{Linearised equations}

In order to linearise the previous set of equations we express every magnitude $\Theta$ as $\Theta= \Theta_0 + \delta \Theta$, where $\Theta_0$ and $\delta \Theta \ll \Theta_0$ are, respectively, the equilibrium value of the magnitude and its perturbation. Then, by considering incompressible perturbations ($\delta \rho_n =0$, we get the linearised equations for momentum and mass conservation:
\begin{equation}
\rho_n \frac{\partial (\delta v_{ni})}{\partial t}= - \frac{\partial (\delta p_n + \rho_n \delta \varphi_n)}{\partial x_i} + \frac{\partial (S_{ik}^{(n)}+T_{ik}^{(n)})}{\partial x_k}
\ ,\label{9}
\end{equation}
\begin{equation}
\frac{\partial (\delta v_{ni})}{\partial x_i} =0  
\ ,\label{10}
\end{equation}
where for simplicity we have defined $\delta \sigma_{ik}^{(n)} \equiv S_{ik}^{(n)}$, $\delta M_{ik}^{(n)} \equiv T_{ik}^{(n)}$, and they are given, respectively, by the following relationships:
\begin{equation}
\frac{\partial S_{ik}^{(n)}}{\partial t} = G_n \left[\frac{\partial (\delta v_{ni})}{\partial x_k} +\frac{\partial (\delta v_{nk})}{\partial x_i} \right]
\ ,\label{11}
\end{equation}
\begin{equation}
T_{ik}^{(n)}= - \frac{B_{nj} \delta B_{nj}}{\mu_n} \delta_{ik} + \frac{B_{ni} \delta B_{nk}}{\mu_n} +\frac{B_{nk} \delta B_{ni}}{\mu_n}
\ ,\label{12}
\end{equation}
where the magnetic field perturbation turns out from (\ref{8}) or (\ref{8a}):
\begin{equation}
\frac{\partial (\delta B_{ni})}{\partial t}= B_{nk} \frac{\partial (\delta v_{ni})}{\partial x_k}
\ .\label{13}
\end{equation}
For the present case (\ref{12}) can be rewritten by taking into account that $B_{nj}=B_n \delta_{jx}$:
\begin{equation}
T_{ik}^{(n)}= \frac{B_n}{\mu_n}(- \delta B_{nx} \delta_{ik} +\delta B_{nk} \delta_{ix} +\delta B_{ni} \delta_{kx} )
\ ,\label{12a}
\end{equation}

To obtain suitable equations for the description of the perturbed velocity field, we use the Helmholtz's decomposition (Lamb 1945, Eringen \& Suhubi 1975, Menikoff et al. 1978, Thorne \& Blandford 2017) whereby the velocity field can be written as the sum of an irrotational part plus a solenoidal part, in terms of the scalar potential $\phi_n$ and the vector potential $\boldsymbol{\psi_n}$, which for the two-dimensional perturbation we consider here, will be written as $\boldsymbol{\psi_n}= \psi_n \boldsymbol{e_z}$  ($\boldsymbol{e_z}$ is the unitary vector in the direction of the $z$-axis). Therefore, we have
\begin{equation}
\delta \boldsymbol{v_n}= \boldsymbol{\nabla}\phi_n + \boldsymbol{\nabla \times}(\psi_n \boldsymbol{e_z})
\ .\label{14}
\end{equation}
By introducing (\ref{14}) into (\ref{9}) and (\ref{10}), we get:
\begin{eqnarray}
 \boldsymbol{\nabla}  \left\{ \frac{\partial}{\partial t} \left(\rho_n \frac{\partial \phi_n}{\partial t} +\delta p_n + \rho_n \delta \varphi_n  \right) + \frac{B_{n}^{2}}{\mu_n}\left[\frac{\partial (\delta v_{nx})}{\partial x} - \frac{\partial^2 \phi_n}{\partial x^2} \right] \right\}  \ \ \ \ \ \  \ \ \ \ \ \  \ \ \ \ \ \  
  \nonumber\\
 + 
 \boldsymbol{\nabla \times} \left[\left(\rho_n \frac{\partial^2 \psi_n}{\partial t^2}- G_n \nabla^2 \psi_n - \frac{B_{n}^{2}}{\mu_n}\frac{\partial^2 \psi_n}{\partial x^2} \right)\boldsymbol{e_z}\right]=0 
. \label{15}
\end{eqnarray}
\begin{equation}
\nabla^2 \phi_n =0
\ .\label{16}
\end{equation}

In our present problem, it is $G_1=G_3=0$, $G_2\equiv G$, and $\boldsymbol{B_2}=0$, so that the second term of (\ref{15}) is equal to zero. Then, by using the so-called Bernoulli gauge (Lamb 1945, Menikoff et al. 1978, Thorne \& Blandford 2017), we can choose $\phi_n$ and $\psi_n$ as solutions of the following system of equations:
\refstepcounter{equation}
$$
\gamma \phi_n +\frac{\delta p_n}{\rho_n} + \delta \varphi_n =0 , \quad
\nabla^2 \phi_n=0 \ ,
  \eqno{(\theequation{\mathit{a},\mathit{b}})}  \label{17}
$$
\refstepcounter{equation}
$$
\psi_1=\psi_3=0, \quad
\gamma^2 \psi_2=\frac{G}{\rho_2} \nabla^2 \psi_2 \ ,
  \eqno{(\theequation{\mathit{a},\mathit{b}})} \label{18}
$$
where $\delta \varphi_n = -g \delta v_{ny}/\gamma$, $\gamma$ is the asymptotic instability growth rate and we have taken two-dimensional perturbations of the form: 
\begin{equation}
\phi_n \propto e^{(\gamma t + qy)} \sin kx
\ ,\label{19}
\end{equation}
\begin{equation}
\psi_2 \propto e^{(\gamma t + q'y)} \cos kx
\  . \label{20}
\end{equation}
In addition, consistently with (\ref{17}$b$) and (\ref{18}$b$) it is:
\refstepcounter{equation}
$$
q= \pm k \ , \quad
q' = \pm \lambda \ , \quad
\lambda = \sqrt{k^2 +\frac{\gamma^2 \rho_2}{G}} .
  \eqno{(\theequation{\mathit{a},\mathit{b},\mathit{c}})} \label{21}
$$
On the other hand, perturbations of the deviatoric stress tensor $S_{ik}^{(2)}$, of the magnetic stress tensors $T_{ik}^{(n)}$, and of the magnetic fields $\delta B_{ni}$ are given by (\ref{11}),  (\ref{12}) and (\ref{13}), respectively. Similarly, the velocity field is obtained from (\ref{14}):
\refstepcounter{equation}
$$
\delta v_{nx}= \frac{\partial \phi_n}{\partial x} + \frac{\partial \psi_n}{\partial y}, \quad
\delta v_{ny}= \frac{\partial \phi_n}{\partial y} - \frac{\partial \psi_n}{\partial x} \ .
  \eqno{(\theequation{\mathit{a},\mathit{b}})} \label{22}
$$

On the other hand, the linearised boundary conditions at each interface are:
\begin{equation}
\| -\delta p_n \delta_{ik} +S_{ik}^{(n)}n_k+ T_{ik}^{(n)}n_k \| =0 
\ ,\label{22a}
\end{equation}
\begin{equation}
\| -\delta v_{ny} \| =0 
\ ,\label{22b}
\end{equation}
(with $i= y$ or $x$) where $\|Q\|=Q(y_0+0^+)-Q(y_0-0^+)$ [with $y_0=\xi_a$ or $y_0=-h+\xi_b$, where $0^+ \rightarrow 0$, and $\xi_a=\xi_a(x,t)$ and $\xi_b=\xi_b(x,t)$ are the perturbation amplitudes of the bottom and top interfaces, respectively], and we have to take into account that both, $S_{ik}^{(n)}$ and $T_{ik}^{(n)}$, are symmetric tensors. By taking into account that in the linear regime we are considering $n_x= \partial \xi_{a,b}/\partial x \sim k\xi_{a,b} <<1$ and $n_y \sim 1$, we can re-write (\ref{22a}) as follows:
\begin{equation}
\| -\delta p_n \delta_{yi} +S_{yi}^{(n)}+ T_{yi}^{(n)} \| =0 
\ ,\label{22c}
\end{equation}

Therefore, the required boundary conditions read:
\refstepcounter{equation}
$$
\delta v_{1y}(0)=\delta v_{2y}(0)=\gamma \xi_a , \quad
\delta v_{2y}(-h)=\delta v_{3y}(-h)= \gamma\xi_b  \ ,
  \eqno{(\theequation{\mathit{a},\mathit{b}})}  \label{54}
$$
\refstepcounter{equation}
$$
-\delta p_1(0)+T_{yy}^{(1)}(0)= -\delta p_2(0)+S_{yy}^{(2)}(0) , \quad
-\delta p_2(-h)+S_{yy}^{(2)}(-h)= T_{yy}^{(3)}(-h) \ ,
  \eqno{(\theequation{\mathit{a},\mathit{b}})}  \label{55}
$$
\refstepcounter{equation}
$$
T_{xy}^{(1)}(0)=S_{xy}^{(2)}(0) , \quad
S_{xy}^{(2)}(-h)=T_{xy}^{(3)}(-h) \ ,
  \eqno{(\theequation{\mathit{a},\mathit{b}})}  \label{56}
$$
where $S_{ik}^{(2)}$ is obtained from (\ref{11}), $T_{ik}^{(n)}$ is obtained from (\ref{12}) and (\ref{13}), and $\delta p_n$ is given by (\ref{17}$a$):
\begin{equation}
- \delta p_n=-\rho_n g \frac{\delta v_{ny}}{\gamma} + \rho_n \gamma \phi_n
\ .\label{28}
\end{equation}

Before using these equations for solving the problem presented in figure 1, let us first to retrieve the results by Lau et al. (2011) for the MRT instability in ideal media ($G=0$), and to consider also the case of the MRT instability in two semi-infinite ($h \rightarrow \infty$) media studied by Sun \& Piriz (2014) in which the heaviest one is an elastic medium. Later, these cases can be used for comparisons with the present problem and to aid the physical interpretations.

\section{Brief overview of some previous relevant results}

\subsection{Ideal fluid slab atop a lighter semi-infinite ideal fluid}

 In this case $\psi_2=0$, and the velocity field is derived from the solution of the Laplace's equation (\ref{16}). The resulting velocity potentials $\phi_n$ in the corresponding regions are conveniently written in the following form:
\begin{equation}
\phi_1 = a_{M1}e^{-ky} e^{\gamma t} \sin kx \ , \ (y \geq 0)
\ ,\label{23}
\end{equation}
\begin{equation}
\phi_2 = \frac{a_{M2} \cosh ky + b_{M2} \cosh k(h+y)}{\sinh kh} e^{\gamma t} \sin kx \ , \ (-h \leq y \leq 0)
\ ,\label{24}
\end{equation}
\begin{equation}
\phi_3 = a_{M3}e^{k(h+y)} e^{\gamma t} \sin kx \ , \ (y \leq -h)
\ ,\label{25}
\end{equation}
where the constants $a_{Mn}$, and $b_{M2}$ will be determined together with the instability growth rate $\gamma$ from the boundary conditions at $y=0$ and $y=-h$ given by (\ref{54}) to (\ref{56}). In this case, since the heavier medium ($n=2$) is also ideal, we have  $S_{ik}^{(2)}=0$, and those boundary conditions read as follows:
\refstepcounter{equation}
$$
\delta v_{1y}(0)=\delta v_{2y}(0) \ , \quad
\delta v_{2y}(-h)=\delta v_{3y}(-h)  ,
  \eqno{(\theequation{\mathit{a},\mathit{b}})} \label{26}
$$
\refstepcounter{equation}
$$
- \delta  p_{1}(0)+ T_{yy}^{(1)}(0)=-\delta p_{2}(0) \ , \quad
- \delta  p_{2}(-h)= - \delta  p_{3}(-h)+ T_{yy}^{(3)}(-h)  ,
  \eqno{(\theequation{\mathit{a},\mathit{b}})} \label{27}
$$
where the perturbations of the magnetic tensor are given by (\ref{12}) and (\ref{13}):
\begin{equation}
T_{yy}^{(n)}= \frac{B_{n}^{2} k^2}{\gamma \mu_n} a_{Mn}
\ .\label{29}
\end{equation}
Then, (\ref{26}) and (\ref{27}) produce the following set of equations:
\refstepcounter{equation}
$$
a_{M1}=-b_{M2} \ , \quad
a_{M2}=-a_{M3}  ,
  \eqno{(\theequation{\mathit{a},\mathit{b}})} \label{30}
$$
\begin{equation}
\gamma \rho_2 \left(\frac{a_{M2}}{\sinh kh} + b_{M2} \coth kh   \right)- \frac{\rho_2  kg}{\gamma} b_{M2}= \rho_1\left( \gamma+ \frac{kg}{\gamma} \right) a_{M1}+ \frac{B_{1}^{2} k^2}{\gamma \mu_1} a_{M1}
\ , \label{31}
\end{equation}
\begin{equation}
\gamma \rho_2 \left(a_{M2} \coth kh + \frac{b_{M2}}{\sinh kh}  \right)+ \frac{\rho_2  kg}{\gamma} a_{M2}=  \frac{B_{3}^{2} k^2}{\gamma \mu_3} a_{M3}
\ .\label{32}
\end{equation}

The solution of this system yields the dispersion relation for the growth rate $\gamma$:
\begin{eqnarray}
\gamma^4 \left(1+\frac{\rho_1}{\rho_2} \coth kh \right)  \ \ \ \ \ \  \ \ \ \ \ \  \ \ \ \ \ \  \ \ \ \ \ \   \ \ \ \ \ \  \ \ \ \ \ \  \ \ \ \ \ \  \ \ \ \ \ \  \ \ \ \ \ \  \ \ \ \ \ \  \ \ \ \ \ \  \ \ \ \ \ \     
  \nonumber\\
  +
   \gamma^2 \left\{ \frac{2 k^2}{\rho_2}(G_{M1}+G_{M3}) \coth kh +\frac{\rho_1}{\rho_2} \left[ \frac{2k^2}{\rho_2}G_{M3} +kg(1+\coth kh) \right]  \right\}    
   \nonumber\\
-
 \left[ \left(1-\frac{\rho_1}{\rho_2}\right)kg-\frac{2k^2}{\rho_2}G_{M1} \right]       \left(kg+ \frac{2k^2}{\rho_2}G_{M3} \right)=0 
, \label{33}
\end{eqnarray}
where we have used the following definition:
\begin{equation}
G_{Mn}\equiv \frac{B_{n}^{2}}{2 \mu_n}
\ .\label{34}
\end{equation}
It may be worth to notice that for $G_{M1}=G_{M3}=0$, (\ref{33}) yields the growth rate obtained by Mikaelian (1982) and Goncharov et al. (2000) for the case of an ideal fluid slab with no magnetic fields:
\begin{equation}
\gamma = \sqrt{\frac{(\rho_2-\rho_1)kg}{\rho_2+ \rho_1 \coth kh}}
\ .\label{35}
\end{equation}
Later, this result will be useful for comparisons.

\begin{figure} 
\centering
\begin{center}
\vspace*{-10 mm}
\includegraphics[width= 1.0 \textwidth, clip]{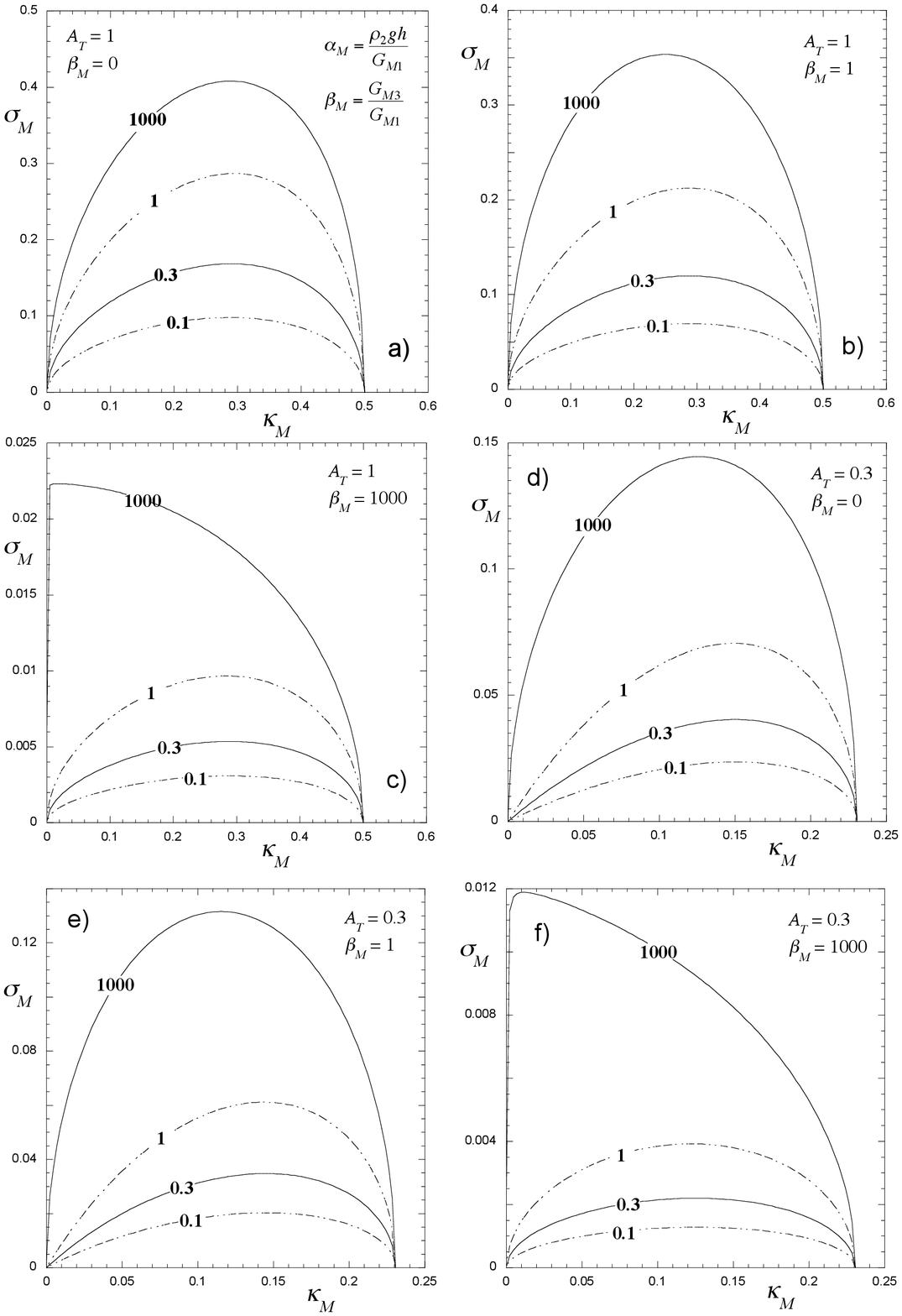}
\end{center}
\vspace*{-5 mm}
\caption{Ideal MRT instability. Dimensionless growth rate $\sigma_M= \gamma/\sqrt{k_{0M}g}$ as a function of the dimensionless wavenumber $\kappa_M = k/k_{0M}$ ($k_{0M}=\rho_2g/G_{M1}$, and $G_{Mn}=B_{n}^{2}/(2 \mu_n)$), for two Atwood numbers ($A_T=1$ and $A_T=0.3$), and for three typical cases in which respectively, the bottom magnetic field dominates [(a) and (d)], both fields are equal [(b) and (e)], and the top magnetic field dominates [(c) and (f)], and for different values of the dimensionless thickness $\alpha_M$ (indicated by the labels on the curves). }
\label{Figure-1}
\end{figure}

For the analysis of the dispersion relation (\ref{33}) it is convenient to introduce the following dimensionless magnitudes: 
\refstepcounter{equation}
$$
\kappa_M = \frac{k}{k_{0M}} \ , \quad
\sigma_M = \frac{\gamma}{\sqrt{k_{0M}g}} \ , \quad
k_{0M}=\frac{\rho_2 g}{G_{M1}} .
  \eqno{(\theequation{\mathit{a},\mathit{b},\mathit{c}})} \label{36}
$$
Then, we get:
\begin{eqnarray}
\sigma_{M}^{4} \left(1+\frac{1-A_T}{1+A_T} \coth \alpha_M \kappa_M  \right)  \ \ \ \ \ \  \ \ \ \ \ \  \ \ \ \ \ \  \ \ \ \ \ \   \ \ \ \ \ \  \ \ \ \ \ \  \ \ \ \ \ \  \ \ \ \ \ \  \ \ \ \ \ \  \ \ \ \ \ \      
  \nonumber\\
  +
\sigma_{M}^{2} \left\{ 2(1+\beta_M)\kappa_{M}^{2} \coth \alpha_M \kappa_M +\frac{1-A_T}{1+A_T} \left[2 \beta_M \kappa_{M}^{2} +\kappa_M(1+\coth  \alpha_M \kappa_M) \right]  \right\}    
   \nonumber\\
-
2 \kappa_{M}^{2} \left( \frac{A_T}{1+A_T}- \kappa_M \right) \left(1+2\beta_M \kappa_M \right)=0  
, \ \ \ \ \ \ \label{37}
\end{eqnarray}
where
\refstepcounter{equation}
$$
\beta_M=\frac{G_{M3}}{G_{M1}} \ , \quad
\alpha_M= k_{0M}h \ , \quad
A_T=\frac{\rho_2-\rho_1}{\rho_2+\rho_1} ,
  \eqno{(\theequation{\mathit{a},\mathit{b},\mathit{c}})} \label{38}
$$
and $A_T$ is the Atwood number. We have presented the dispersion relation (\ref{37}) in a somewhat different manner from Lau et al. (2011), that is more suitable to our present purposes. The dimensionless growth rate $\sigma_M$ is shown in figure 2 for two values of the Atwood number ($A_T=1$ and $0.3$), for three values of the ratio $\beta_M=G_{M3}/G_{M1}$ between the magnetic pressures above and below the dense layer, and for different values of its dimensionless thickness $\alpha_M$ indicated by the labels on the curves.

Three features have to be noticed that will be worth comparing later on with the results involving an elastic slab in \S 4. First, the cut-off wavenumber $k_{Mc}$ is independent of the slab thickness and it is determined only by the magnetic pressure   $G_{M1}=B_{1}^{2}/(2\mu_1)$ acting on the bottom surface of the slab:
\begin{equation}
k_{Mc} =\frac{A_T}{1+A_T}\frac{\rho_2 g}{G_{M1}}
\ .\label{39}
\end{equation}
Secondly, the maximum growth rate is smaller for the thinner slabs, and this is true for all the values of $A_T$ whereas, as shown by (\ref{35}), when no magnetic fields are present the thinner slabs are more stable only if $A_T \neq 1$. As it was discussed by Piriz, Piriz \& Tahir (2018), the latter behaviour is explained by the fact that the fluid in the region $y\geq 0$ exerts a force (per unitary surface) $\rho_1g\xi_a$ which opposes the slab fall, and it depends only on the bottom surface deformation $\xi_a$. Instead, the force $\rho_2g \Delta h$ driving the slab fall is given by the slab-thickness change $\Delta h=\xi_a-\xi_b$. Therefore, since $\Delta h$ is smaller for thinner slabs, and the same force  $\rho_1g\xi_a$ is available to support any slab, the thicker ones are less stable. For the same reason, no reduction of the growth rate occurs for $A_T=1$, when no fluid is present below the slab for supporting it.

However, the fact that in presence of magnetic fields the thinner slabs turn out to be more stable even for $A_T=1$, can be explained in a similar manner. In fact, uniform magnetic fields which are known to act as a surface tension (Chandrasekhar 1961), resist the slab deformation in an extent that depends on the local deformations $\xi_a$ at $y=0$,  and $\xi_b$ at $y=-h$  [in (\ref{29}) it is $a_{M1}/\gamma= \xi_a$ and  $a_{M3}/\gamma= \xi_b$]. Instead, the weight increase of the slab is once again determined by $\Delta h = \xi_a- \xi_b$, which is smaller for thinner slabs. The stabilising effect of the magnetic field is then more effective for the thinner slabs for any $A_T$.

Finally, we note that for a given value of $A_T$, the magnetic field atop the layer also acts to reduce the maximum growth rate. This means that the presence of magnetic fields at any side of the slab has a stabilising effect.

\subsection{MRT in semi-infinite media. An elastic medium atop an ideal fluid}

In this case $h \rightarrow -\infty$, and $\psi_2 \neq 0$ is given by (\ref{18}). Then, the potential functions that determine the velocity field read as follows:
\begin{equation}
\phi_1 = a_{s1}e^{-ky} e^{\gamma t} \sin kx \ , \ (y \geq 0)
\ ,\label{40}
\end{equation}
\begin{equation}
\phi_2 = a_{s2}e^{ky} e^{\gamma t} \sin kx \ , \ (y \leq 0)
\ ,\label{41}
\end{equation}
\begin{equation}
\psi_2 = c_{s2}e^{\lambda y} e^{\gamma t} \cos kx \ , \ (y \leq 0)
\ ,\label{42}
\end{equation}
where $\lambda$ is given by (\ref{21}$c$). 

In this case, the boundary conditions must be imposed only at $y=0$. Therefore, (\ref{54}) to (\ref{56}) read now as follows:
\begin{equation}
\delta v_{1y}(0)=\delta v_{2y}(0)
\ ,\label{43}
\end{equation}
\begin{equation}
-\delta p_1(0)+T_{yy}^{(1)}(0)= -\delta p_2(0)+S_{yy}^{(2)}(0)
\ ,\label{44}
\end{equation}
\begin{equation}
T_{xy}^{(1)}(0)=S_{xy}^{(2)}(0)
\ .\label{45}
\end{equation}

These equations yield the following system for obtaining the constants $a_{sn}$ and $c_{s2}$, and the growth rate $\gamma$.
\refstepcounter{equation}
$$
-a_{s1}=a_{s2}+c_{s2} , \quad
a_{s2} =- \frac{\left[(\lambda^2 +k^2)G-2k^2 G_{M1} \right]} {2 k^2 (G-G_{M1})} c_{s2} \ ,
  \eqno{(\theequation{\mathit{a},\mathit{b}})}  \label{46}
$$
\begin{equation}
\rho_1(\gamma +\frac{kg}{\gamma}) a_{s1}+\frac{2G_{M1}k^2}{\gamma} a_{s1}= \rho_2 \gamma a_{s2}-\frac{kg}{\gamma} \rho_2 (a_{s2}+c_{s2})+\frac{2G}{\gamma}(k^2 a_{s2}+\lambda k c_{s2})
\ ,\label{47}
\end{equation}
From (\ref{46}) and (\ref{47}) we can find the dispersion relation for the growth rate:
\begin{equation}
\gamma^2-A_Tkg= \frac{1+A_T}{2} \frac{4k^2G}{\rho_2}\left[-1+\frac{kG}{\rho_2 \gamma^2} \left(\sqrt{k^2+\frac{\gamma^2 \rho_2}{G}}-k \right)\left(1-\frac{G_{M1}}{G}\right) \right]
\ .\label{47a}
\end{equation}
Notice that by doing the transformation $ G/\gamma \rho_2 \rightarrow  \nu$ in the previous equation with $G_{M1}=0$ we recover the growth rate given by Chandrasekhar (1961) for the case of a semi-infinite viscous fluid of dynamic viscosity $\nu$ overlaying a semi-infinite ideal fluid (Robinson \& Swegle 1989).

After some straightforward algebra (\ref{47a}) can be re-written in the following suitable form:
\begin{equation}
\gamma^2-A_Tkg=-\frac{1+A_T}{2} \frac{4k^2}{\rho_2}\frac{(\lambda G + kG_{M1})}{\lambda+k}
\ .\label{48}
\end{equation}
In the irrotational approximation considered by Sun \& Piriz (2014) it is $k^2 \gg \gamma^2\rho_2/G$, and then $\lambda \approx k$, so that (\ref{48}) reduces to the irrotational solution.

From (\ref{48}) we can easily see that marginal stability ($\gamma=0$) occurs for the cut-off wavenumber $k_c$:
\begin{equation}
k_c=\frac{A_T}{1+A_T} \frac{\rho_2 g}{G+G_{M1}}
\ .\label{49}
\end{equation}
As it was noticed by Sun \& Piriz (2014), in this case magnetic pressure acts practically in the same manner than elasticity to enhance the stability of the interface, and both stabilising effects are added up.

To represent graphically the dispersion relation for this case it is more convenient to write (\ref{47a}) in dimensionless form by using the following definitions:
\refstepcounter{equation}
$$
\kappa=\frac{k}{k_0} , \quad
\sigma= \frac{\gamma}{\sqrt{k_0g}} , \quad
k_0=\frac{\rho_2g}{G} , \quad
\beta_1= \frac{G_{M1}}{G} \ .
  \eqno{(\theequation{\mathit{a},\mathit{b},\mathit{c}})}  \label{49a}
$$
Then, it turns out:
\begin{equation}
\sigma^2= A_T \kappa + 2(1+A_T) \kappa^2 \left\{ - 1 + \frac{\kappa}{\sigma^2}\left[\sqrt{\kappa^2+ \sigma^2}-\kappa \right](1-\beta_1)\right\}
\ .\label{49b}
\end{equation}
This expression is represented in figure 3 for two values of the Atwood number ($A_T=1$ and $0.3$) and for three different values of the ratio $\beta_1$ ($0$, $0.5$, and $1.5$).
\begin{figure} 
\centering
\begin{center}
\vspace*{-10 mm}
\includegraphics[width= 1.0 \textwidth, clip]{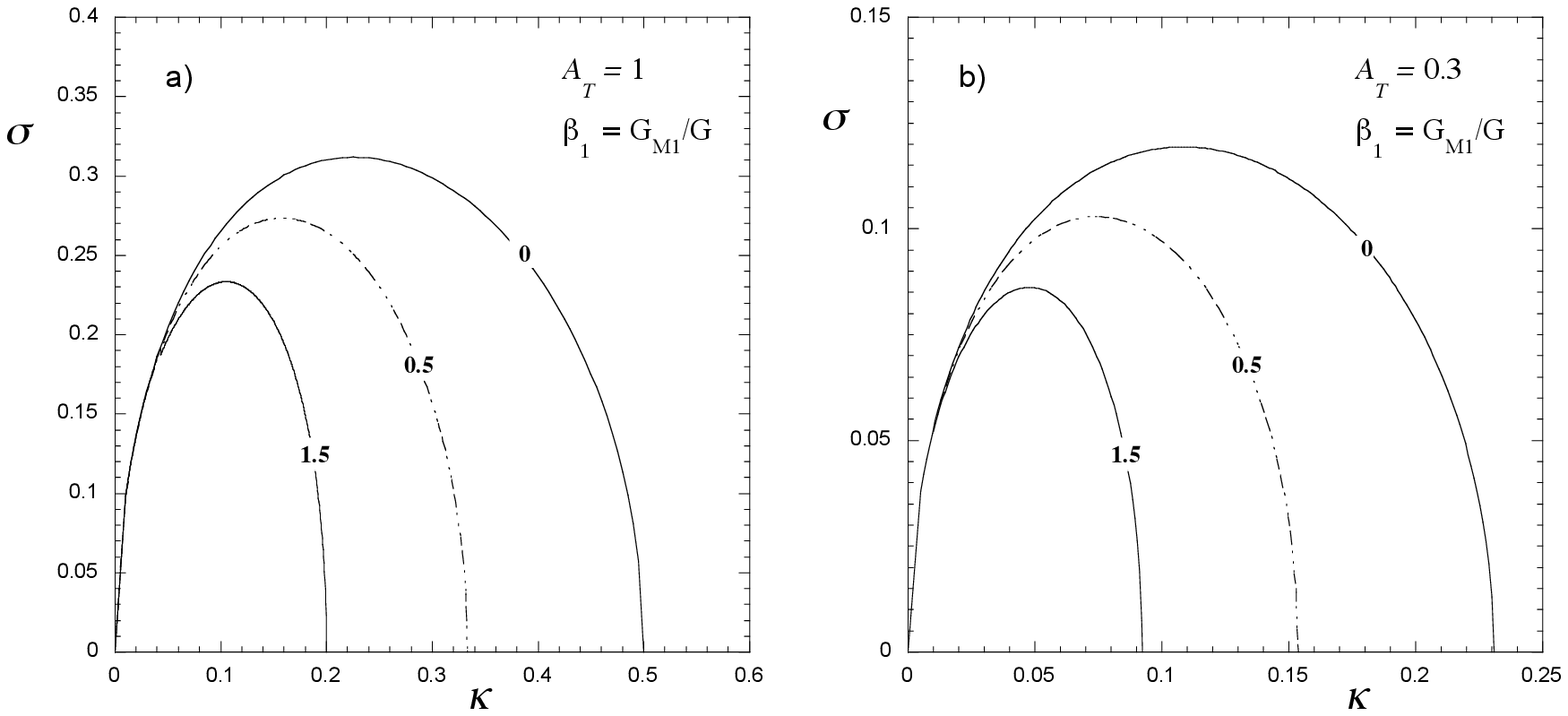}
\end{center}
\vspace*{-120 mm}
\caption{MRTI in semi-infinite media with an elastic medium atop an ideal fluid. Dimensionless growth rate $\sigma= \gamma/\sqrt{k_{0}g}$ as a function of the dimensionless wavenumber $\kappa = k/k_{0}$ ($k_{0}=\rho_2g/G$) for three values of the ratio $\beta_1= G_{M1}/G$ [$G_{M1}=B_{1}^{2}/(2 \mu_1)$] indicated by the labels on the curves, and 
for two Atwood numbers: a) $A_T=1$ and b) $A_T=0.3$. }
\label{Figure-1}
\end{figure}

\section{Elastic slab atop an ideal fluid in presence of magnetic fields}

\subsection{Dispersion relation}

This is the case schematically represented in figure 1 in which the regions $y\leq -h$ and $y\geq 0$ are filled with respective uniform magnetic fields, and the elastic slab in between ($-h\leq y \leq 0$) is a Hookean medium.  Then $\psi_2 \neq 0$ and the potential functions are conveniently written in the following form:
\begin{equation}
\phi_1 = a_{1}e^{-ky} e^{\gamma t} \sin kx \ , \ (y \geq 0)
\ ,\label{50}
\end{equation}
\begin{equation}
\phi_3 = a_{3}e^{k(h+y)} e^{\gamma t} \sin kx \ , \ (y \leq -h)
\ ,\label{51}
\end{equation}
\begin{equation}
\phi_2 = \frac{a_{2} \cosh ky + b_{2} \cosh k(h+y)}{\sinh kh} e^{\gamma t} \sin kx \ , \ (-h \leq y \leq 0)
\ ,\label{52}
\end{equation}
\begin{equation}
\psi_2 = \frac{c_{2} \sinh \lambda y + d_{2} \sinh \lambda (h+y)}{\sinh \lambda h} e^{\gamma t} \cos kx \ , \ (-h \leq y \leq 0)
\ ,\label{53}
\end{equation}
and the velocity field is calculated from (\ref{22}).

The required boundary conditions are those ones given by  (\ref{54}) to (\ref{56}). These six boundary conditions produce the following system of linear equations:
\refstepcounter{equation}
$$
-a_1=b_2 +d_2 , \quad
-a_3=a_2+c_2 \ ,
  \eqno{(\theequation{\mathit{a},\mathit{b}})}  \label{55}
$$
\begin{equation}
a_1= - \frac{G\left[2k^2b_2+(\lambda^2 +k^2)d_2 \right]}{2k^2G_{M1}}
\ ,\label{57}
\end{equation}
\begin{equation}
a_3= - \frac{G\left[2k^2a_2 +(\lambda^2 +k^2)c_2\right]}{2k^2G_{M3}}
\ ,\label{58}
\end{equation}
\begin{eqnarray}
\gamma \rho_2  \left(\frac{a_2}{\sinh kh}+b_2 \coth kh \right)  \ \ \ \ \ \  \ \ \ \ \ \  \ \ \ \ \ \  \ \ \ \ \ \   \ \ \ \ \ \  \ \ \ \ \ \  \ \ \ \ \ \  \ \ \ \ \ \  \ \ \ \ \ \  \ \ \ \ \ \      
  \nonumber\\
  +
\frac{2kG}{\gamma} \left[k\left(\frac{a_2}{\sinh kh}+b_2 \coth kh \right) + \lambda \left(\frac{c_2}{\sinh \lambda h}+d_2 \coth \lambda h \right) \right]  \ \ \ \ \ \  
   \nonumber\\
-
\frac{\rho_2kg}{\gamma} (b_2+d_2)= \rho_1\left(\gamma +\frac{kg}{\gamma} \right) a_1 + \frac{2k^2G_{M1}}{\gamma} a_1
, \label{59}
\end{eqnarray}
\begin{eqnarray}
\gamma \rho_2 \left(a_2 \coth kh+ \frac{b_2}{\sinh kh} \right)  \ \ \ \ \ \  \ \ \ \ \ \  \ \ \ \ \ \  \ \ \ \ \ \   \ \ \ \ \ \  \ \ \ \ \ \  \ \ \ \ \ \  \ \ \ \ \ \  \ \ \ \ \ \  \ \ \ \ \ \      
  \nonumber\\
  +
\frac{2kG}{\gamma} \left[k\left(a_2 \coth kh+\frac{b_2}{\sinh kh} \right) + \lambda \left(c_2 \coth \lambda h+ \frac{d_2}{\sinh \lambda h}  \right) \right]  \ \ \ \ \ \  
   \nonumber\\
+
\frac{\rho_2kg}{\gamma} (a_2+c_2) = \frac{2k^2G_{M3}}{\gamma} a_3
. \label{60}
\end{eqnarray}

After some algebra the previous six-equations system can be reduced to the following two-equations system:
\begin{equation}
a_2 (A_1+E_1) + b_2\left[C_1-B+\frac{\rho_1}{\rho_2}\left(B +\frac{\gamma^2 \rho_2}{\gamma} \right)\right]=0
\ ,\label{61}
\end{equation}
\begin{equation}
a_2(C_3 +B)+b_2 (A_3-E_3)=0
\ ,\label{62}
\end{equation}
where
\begin{equation}
A_n=(1-\beta_n)A + \beta_n(\lambda^2+k^2) \mathrm{csch} kh
\ ,\label{63}
\end{equation}
\begin{equation}
C_n=(1-\beta_n)C + \beta_n\left[(\lambda^2+k^2) \mathrm{coth} kh+2k^2\right]
\ ,\label{64}
\end{equation}
\begin{equation}
E_n= \frac{4k^3 \lambda(\lambda^2+k^2-2k^2 \beta_n)}{\lambda^2-k^2} E_0
\ ,\label{64a}
\end{equation}
\begin{equation}
 B=\frac{\rho_2kg}{G} 
\ ,\label{65}
\end{equation}
and we have used the following definitions:
\refstepcounter{equation}
$$
E_0 =\left( \frac{1-\beta_1}{\lambda^2+k^2-2k^2 \beta_1}-\frac{1-\beta_3}{\lambda^2+k^2-2k^2 \beta_3}\right)\mathrm{coth} \lambda h \ ; \  \ \beta_n =\frac{B_{n}^{2}}{2\mu_nG}
  \eqno{(\theequation{\mathit{a},\mathit{b}})} \label{66}
$$
\begin{equation}
A=\frac{(\lambda^2+k^2)^2\mathrm{csch} kh -4k^3 \lambda \mathrm{csch} \lambda h}{\lambda^2-k^2}
\ ,\label{67}
\end{equation}
\begin{equation}
C=\frac{(\lambda^2+k^2)^2\mathrm{coth} kh -4k^3 \lambda \mathrm{coth} \lambda h}{\lambda^2-k^2}
\ . \label{68}
\end{equation}
We note that $E_n=0$ when $\beta_1=\beta_3$ and also when $\gamma=0$. In addition, when $\beta_1\gg 1$ or $\beta_3 \gg 1$ , it is $E_n/A_n < \gamma^2 \rho_2/(2k^2 G) <1$. Therefore, $E_n$ will have some slightly effect only on the maximum growth rate for the latter extreme cases and can be neglected in all the situations of interest.
Then, by neglecting hereafter such terms, the dispersion relation turns out from the solution of the system (\ref{61}) and (\ref{62}):
\begin{equation}
C_1C_3-A_1A_3=B^2-\frac{\rho_1}{\rho_2}\left(B+\frac{\gamma^2 \rho_2}{G} \right)(B+C_3)-B(C_1-C_3)
\ .\label{69}
\end{equation}

\subsection{Marginal stability conditions}

Before proceeding with the calculation of the instability growth rate $\gamma$, it is very useful to study the conditions for marginal stability by solving (\ref{69}) for the case with $\gamma(k=k_c)=0$ ($k_c$ is a cut-off wavenumber). Then, and by using the L'H\^{o}pital's rule, we find that for $\gamma \rightarrow 0$ (\ref{63}) and (\ref{64}) yield:
\begin{equation}
A_n=\frac{2k_{c}^{2} \left[k_ch(1-\beta_n)\cosh k_ch +\sinh k_ch \right]}{(\sinh k_ch)^2}
\ ,\label{70}
\end{equation}
\begin{equation}
C_n=\frac{C'}{h^2}+2 k_{c}^{2} \beta_n
\ ,\label{71}
\end{equation}
where
\refstepcounter{equation}
$$
C_{n}^{'}=\left(\frac{w}{\sinh w}\right)^2\left[2w(1-\beta_n)+\sinh 2w \right] , \quad
w= k_ch \ .
  \eqno{(\theequation{\mathit{a},\mathit{b}})}  \label{72}
$$
Then, (\ref{69}) leads to the following equation for the marginal stability conditions:
\begin{equation}
\frac{2A_T}{1+A_T} \alpha^2 -\alpha H_1(w)-H_2(w)=0
\ ,\label{73}
\end{equation}
where $\alpha=\rho_2gh/G$, and
\begin{equation}
H_1= \frac{(1-A_T)w}{1+A_T}  \left[\frac{2w(1-\beta_3) +\sinh 2w}{(\sinh w)^2} +2\beta_3\right]+2(\beta_1-\beta_3)\left[w-\left(\frac{w}{\sinh w}\right)^2\right]
 ,\label{74}
\end{equation}
\begin{equation}
H_2= 4w^2\left[1-\frac{(1-\beta_1)(1-\beta_3)w^2}{(\sinh w)^2}\right] +4 w^2 \beta_1 \beta_3+2(\beta_1C_{1}^{'} +\beta_3C_{3}^{'}) 
\ ,\label{75}
\end{equation}
where $C_{n}^{'}$ is given by (\ref{72}$a$).

It is worth to notice that in the limit $w \gg 1$, $\alpha \gg 1$ (very thick slabs) (\ref{73}) gives:
\refstepcounter{equation}
$$
\kappa_c=\frac{w}{\alpha}=\frac{A_T}{(1+A_T)(1+\beta_1)}= \frac{k}{k_0} ,  \quad
k_0= \frac{\rho_2g}{G} \ .
  \eqno{(\theequation{\mathit{a},\mathit{b}})}  \label{76}
$$

That is, in the limit of very thick slabs we retrieve the result of \S 3.2 whereby the cut-off wavenumber is determined only by the magnetic field $\boldsymbol{B_1}$ beneath the slab and is not affected by the field $\boldsymbol{B_3}$ atop it.

However, in the opposite limit, $w \ll1$, (\ref{73}) gives place in general to a variety of different behaviours of the marginal stability curves depending on the values of $\beta_n$, and $A_T$. In such a limit, (\ref{74}) and (\ref{75}) reduce to the following forms:
\begin{equation}
H_1(w)\approx \frac{1-A_T}{1+A_T}\left[2(1-\beta_3)+2+2w\beta_3\right]-2(\beta_1-\beta_3)(1-w)
\ .\label{77}
\end{equation}
\begin{equation}
H_2(w)\approx 4w^2\left[1-(1-\beta_1)(1-\beta_3)(1-\frac{w^2}{3}) \right]+4w[\beta_1(2-\beta_3)+\beta_3(2-\beta_1)]
\ .\label{78}
\end{equation}
It is convenient to analyse this limit of $w\ll1$ separately for the most representative cases.
\begin{figure}
\centering
\begin{center}
\vspace*{-2 mm}
\includegraphics[width= 1.0 \textwidth, clip]{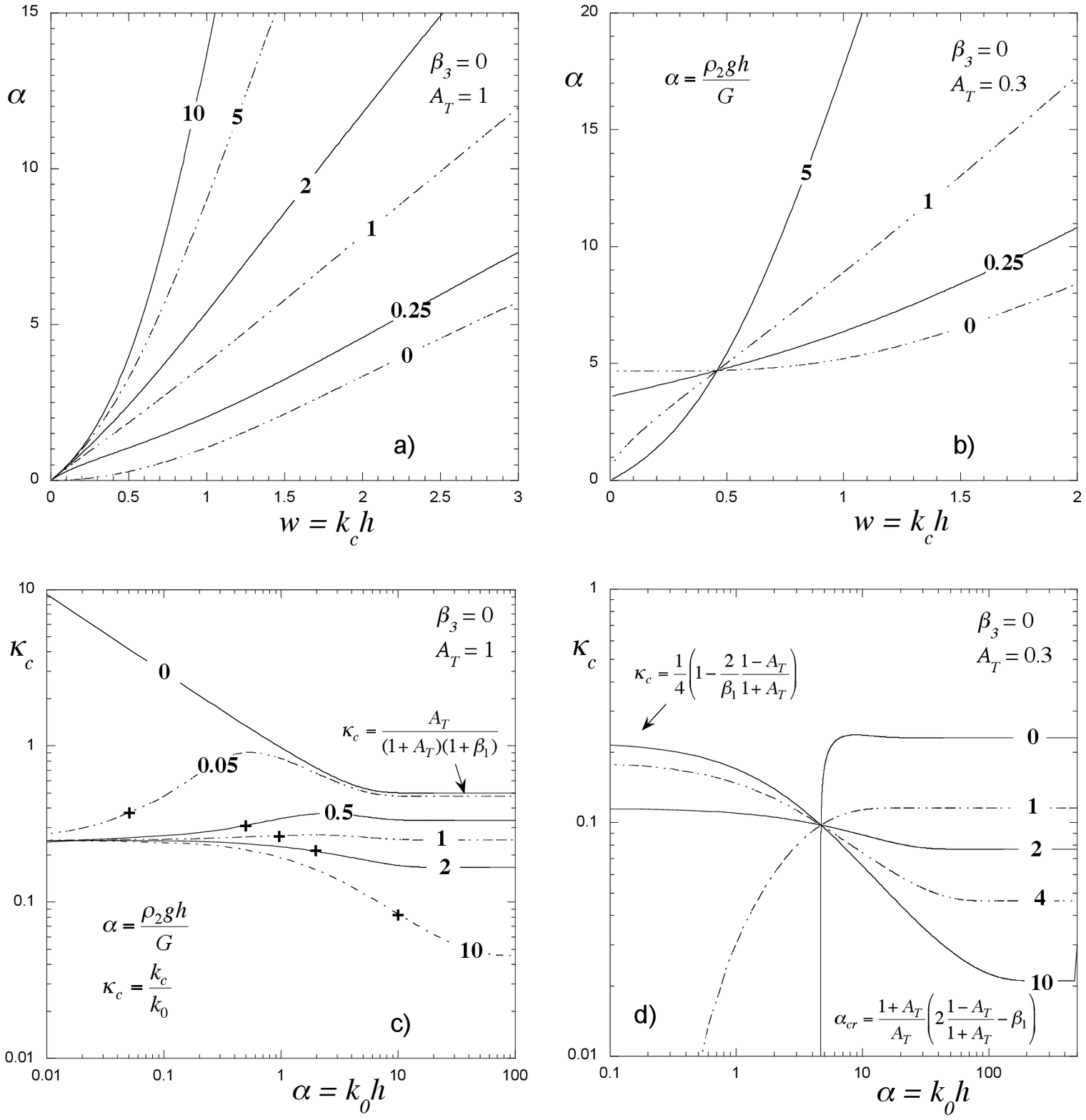}
\end{center}
\vspace*{-55 mm}
\caption{Marginal stability curves for the case with no magnetic field atop the slab, $\beta_3=0$ [$\beta_n=B_{n}^{2}/(2\mu_nG)$], for two Atwood numbers ($A_T=1$ and $A_T=0.3$), and for several values of $\beta_1$ (indicated by the labels on the curves). (a) and (b) Dimensionless slab thickness $\alpha$ as a function of $w=k_ch$. (c) and (d) Dimensionless cut-off wavenumber $\kappa_c=w/\alpha$ as a function of the $\alpha$.}
\label{Figure-1}
\end{figure}

\subsubsection{$\beta_1 \geq 0$, $\beta_3=0$}

In this case, by putting $\beta_3=0$ in (\ref{77}) and (\ref{78}) we find that (\ref{73}) reduces to:
\begin{equation}
\frac{2A_T}{1+A_T}\alpha^2-2\alpha \left[\frac{2(1-A_T)}{1+A_T}-\beta_1(1-w) \right]= 8w\beta_1+4w^2\left[1-(1-\beta_1)(1-\frac{w^2}{3}) \right]
\ .\label{79}
\end{equation}
Then, for $A_T=1$ and provided that $\beta_1\neq0$, we get ($w \ll 1$, $\alpha \ll 1$):
\begin{equation}
\alpha=4w \ , \  \mathrm{or}   \  \  \  \kappa_c= \frac{w}{\alpha}=\frac{1}{4}
\ .\label{80}
\end{equation}
Instead, when $\beta_1=0$, we recover the results by Bakhrakh et al. (1997), and Plohr \& Sharp (1998) :
\begin{equation}
\kappa_c =\left(\frac{3}{4} \right)^{1/4}\frac{1}{\sqrt{\alpha}}
\ .\label{81}
\end{equation}

When $A_T \neq1$ we can see from (\ref{79}) that two different behaviours are obtained depending on the values of $\beta_1$. In fact, the cut-off wavenumber results to be:
\begin{equation}
\kappa_c =\left(\frac{1}{4} \right)\left(1-\frac{2}{\beta_1}\frac{1-A_T}{1+A_T}\right)
\ ,\label{82}
\end{equation}
provided that $\beta_1 \geq 2(1-A_T)/(1+A_T)$. If not, a critical value $\alpha_{cr}$ exists below which the system is stable for any perturbation wavenumber:
\begin{equation}
\alpha_{cr}=\frac{1+A_T}{A_T}\left(2 \frac{1-A_T}{1+A_T}-\beta_1\right)
\ ,\label{83}
\end{equation}
from which we retrieve the result by Piriz, Piriz \& Tahir (2017b) when $\beta_1=0$ and no magnetic field is present.

The general results for $\beta_3=0$ are represented in figure 4 for $A_T=1$ and $A_T=0.3$. Figures 4(a) and 4(b) show the dimensionless thickness $\alpha$ as a function of $w=k_ch$, and figures 4(c) and 4(d) show the dimensionless cut-off wavenumber $\kappa_c$ as a function of the dimensionless slab thickness $\alpha=\rho_2gh/G$. It is seen that,   in accordance with (\ref{82}) and (\ref{83}) for $w\ll1$ ($\alpha \ll 1$), when $A_T \neq1$ the cut-off wavenumber becomes larger as the dimensionless magnetic pressure beneath the slab increases, and that the slab is completely stable for $\beta_1<2(1-A_T)/(1+A_T)$. When $A_T=1$ the cut-off wavenumber always decreases as $\beta_1$ increases, but the effect becomes progressively weaker for the thinner slabs, in such a way that for very thin slabs it becomes independent of $\beta_1$ and $\kappa_c \rightarrow 1/4$. 

In other words, contrary to the behaviour for $\alpha \gg1$ in which the effects of elasticity and magnetic field are added together for enhancing the stability, for the thinner slabs the presence of the magnetic pressure in the region beneath the slab makes the system less stable than when only one of these, otherwise stabilising mechanisms, is present. This rather unexpected result is less evident for $A_T \sim 1$ [figure 4(c)], but it already indicates that the presence of the magnetic field is not significantly affecting the cut-off wavenumber for the thinner slabs. However, the effect becomes very evident for $A_T <1$ [figure 4(d)]. This means that when the stabilising effect of the magnetic field is present it enters into competition with the stabilising effect of elasticity in such a manner that the former acts in opposition to the latter. 

This competition phenomenon is connected with the fact that the magnetic field stabilising effect is determined by the local strain $k\xi_a$ at $y=0$, while the stabilising effect of the elasticity depends on the total strain which, for relatively thin slabs, is of the order of $(\xi_a-\xi_b)/h$ (Piriz, Piriz \& Tahir 2017a, 2017b). For the very thick slabs the total strain coincides with the local one, and the stabilising effects of the magnetic field add to the ones of the elasticity. However, for the thinner slabs, the total strain is affected by the presence of the magnetic field which acts to reduce the local deformation $\xi_a$. This leads to a reduction of $\xi_a-\xi_b$, which for a given slab thickness $h$, reduces the effectivity of the stabilising effect of elasticity. Such a reduction is stronger for thinner slabs and for higher magnetic pressures.

In addition, elasticity tends to resist the stabilising effect of the magnetic field mitigating the deformation $\xi_a$ of the interface at $y=0$. As a result the system may becomes less stable than when only one of these mechanisms is present. It is seen in figure 4(d), there is some specific value $\alpha* \approx 4.73$ whereby the stabilising effects and the competition between magnetic field and elasticity mutually compensate each other, and $\kappa_c$ results to be independent of $\beta_1$ ($\kappa_c \approx 0.1$). For $\alpha <\alpha*$ the elasticity and the magnetic field act against each other, while in the opposite case they act in the same sense until for $\alpha \gg 1$ they are linearly added up.

Such results can be of concern for some experiments on high energy density physics in which flyer plates are accelerated in such a manner to keep the plates in solid state with the aim to increase the acceleration stability (Lemke et al. 2011, Martin et al. 2012). In such cases, $A_T \approx 1$ and assuming that the plate is driven exclusively by the magnetic pressure, we have $\alpha=\beta_1$. In fact, the equilibrium condition imposes the following relationship at the bottom interface ($y=0$), given by (\ref{4a}):
\begin{equation}
\rho_2gh=p_1 +\frac{B_{1}^{2}}{2\mu_1}-\frac{B_{3}^{2}}{2\mu_3} , \   \mathrm{or} \  \  \  \alpha=\frac{p_1}{G}+\beta_1-\beta_3
\ ,\label{84}
\end{equation}
so that for $p_1=0$ and $\beta_3=0$ it turns out $\alpha=\beta_1$. We have indicated this particular case with crosses in figure 4(c).

On the other hand, the existence of this competition phenomenon is of relevance for the generation of crust-quakes in the strongly magnetised neutron stars known as magnetars. It is to be noted that, although the enormous magnetic fields exist on the surface of magnetars, there are evidences of much stronger magnetic fields beneath the neutron star crust (Cooper \& Kaplan 2010, Ryu et al. 2012, Mereghetti, Pons \& Melator 2015). Besides, it has been shown that in the absence of magnetic fields, the sole effect of the crust elasticity imposes an instability threshold that depends on the magnitude of the density inversion in the neutron star crust (Blaes et al.1990, 1992, Piriz, Piriz \& Tahir 2017b). This density inversion produced by pycnonuclear and electron capture reactions in the crust (Blaes et al.1990, 1992, Mock \& Joss. 1998, Bildsten \& Cummins, 1998) leads to a maximum Atwood number close to $0.02$, and it was shown to be insufficient for overtaking the purely elastic instability threshold (Mock \& Joss. 1998).  The presence of  magnetic fields was not taken into account in those works probably because, on the basis of the current knowledge on thick media, it was assumed that it would further increase the instability threshold. However, the present results show a different scenario in which a magnetic field such that $B_{1}^{2}/(2\mu_1) \sim G$ can completely remove such an instability threshold and lead to the crust instability for any arbitrary small density inversion. Since a reasonably value for the shear modulus of the neutron star crust is $G \sim 10^{18}$ Mbar (Lander et al. 2015), a magnetic field $B_1 \sim 10^{16}$ Gauss, would be sufficient to make the crust unstable for any perturbation. Such a value of $B_1$ is well within the range of values expected for the internal fields in magnetars. We will see later than the presence of a magnetic field on the crust surface ($\beta_3 \neq0$) does not alter this conclusion.

\subsubsection{$\beta_1 = 0$, $\beta_3 \geq 0$}

By putting now $\beta_1=0$ in (\ref{77}) and (\ref{78}) we get for $w \ll1$:
\begin{equation}
H_1(w)\approx 2 \frac{1-A_T}{1+A_T} (2-\beta_3) +2\beta_3(1-w)
\ ,\label{85}
\end{equation}
\begin{equation}
H_2(w)\approx 8\beta_3w 
\ .\label{86}
\end{equation}
Then, from (\ref{73}) we obtain the value of the dimensionless slab thickness $\alpha$ for $w=0$:
\begin{equation}
\alpha(w=0)=\frac{2(1-A_T)}{A_T} + 2\beta_3 
\ .\label{87}
\end{equation}
This is the minimum value of $\alpha$ below which the system is stable for any perturbation wavenumber provided that $d\alpha/dw|_{w=0} \geq0$. Otherwise, there is a minimum value $\alpha_{min} \leq \alpha(w=0)$ that determines the critical value $\alpha_{cr}$ for the instability threshold. It is not difficult to see that
\begin{equation}
\left. \frac{d\alpha}{dw} \right |_{w=0}=\frac{(1+A_T)\beta_3(2-\beta_3)}{A_T\beta_3+1-A_T}
\ ,\label{88}
\end{equation}
so that there exists a minimum value $\alpha_{min}\leq \alpha(w=0)$ provided that $\beta_3 \geq2$.

We show $\alpha(w)$ in figures 5(a) and 5(b) for two different values of the Atwood number and for the several values of $\beta_3$ indicated by the labels on the curves. As it can be seen in figure 5(a), for $A_T=1$ it is $\alpha(w=0)=2\beta_3$, for which an instability threshold exists only for $\beta_3>0$, as indicated by (\ref{87}). And we can also appreciate the appearance of a minimum for $\beta_3 \geq 2$. Instead, for $A_T=0.3$ there is always a threshold for any value of $\beta_3$, and once again, a minimum smaller than $\alpha(w=0)$ appears for $\beta_3\geq2$.
\begin{figure} 
\centering
\begin{center}
\vspace*{-2 mm}
\includegraphics[width= 1.0 \textwidth, clip]{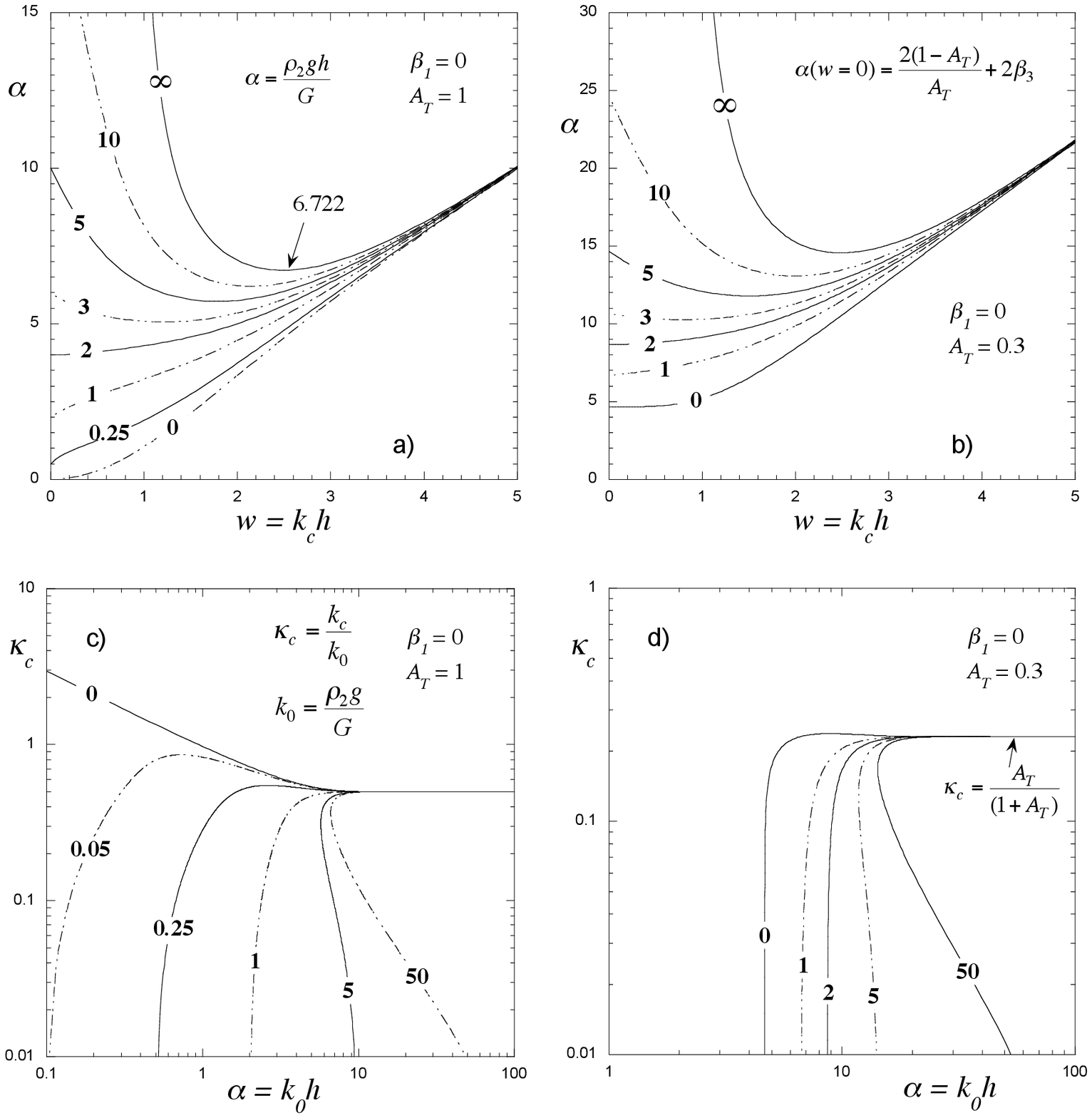}
\end{center}
\vspace*{-55 mm}
\caption{Marginal stability curves for the case with no magnetic field beneath the slab, $\beta_1=0$ [$\beta_n=B_{n}^{2}/(2\mu_nG)$], for two Atwood numbers ($A_T=1$ and $A_T=0.3$), and for several values of $\beta_3$ (indicated by the labels on the curves). (a) and (b) Dimensionless slab thickness $\alpha$ as a function of $w=k_ch$. (c) and (d) Dimensionless cut-off wavenumber $\kappa_c=w/\alpha$ as a function of the $\alpha$.}
\label{Figure-1}
\end{figure}

Figures 5(c) and 5(d) show the same cases as before but for the marginal stability wavenumber $\kappa_c$ as a function of dimensionless thickness $\alpha$. The behaviour is qualitatively the same for any Atwood number except for the fact that, for the purely elastic case ($\beta_3=0$), there is no instability threshold when $A_T=1$ (Plohr \& Sharp 1998, Piriz, Piriz \& Tahir 2017a, 2017b).

Figures 5(a) and 5(b) also show that there are two cut-off wavenumbers for a given value of $\alpha$ when $\beta_3 >2$. This situation resembles the results found by Mora et al. (2014) and Ricobelli \& Ciarletta (2017) for the case of an elastic slab in contact with a rigid surface. We can analytically find the curve $\alpha(w)$ for the limiting case $\beta_3 \rightarrow \infty$:
\begin{equation}
\alpha_{\infty}(w)=\frac{1+A_T}{2A_T}\frac{w(2w^2+2w+\sinh 2w)}{(\sinh w)^2-w}
\ . \label{89}
\end{equation}
This curve has a minimum value for $w_m\approx 2.487$, and the minimum value of $\alpha_{\infty}$ is
\begin{equation}
\alpha_{\infty}(w_m)\approx 6.722 \frac{1+A_T}{2A_T}
\ . \label{90}
\end{equation}

The previous results are similar to the ones obtained for the case of an elastic slab with rigid wall boundary conditions. However, even for a very strong magnetic field, the case involving rigid walls is never retrieved. This is because, although the normal velocity perturbation at $y=-h$, $\delta v_{2y}(-y)\rightarrow 0$ for $\beta_3\rightarrow \infty$, the tangential velocity $\delta v_{2x}(-h)$ remains finite. Instead, in a rigid wall the no-slipping boundary condition imposes that $\delta v_{2x}(-h)$ must also be equal to zero. Nevertheless, the behaviour found here for $\beta_3 >2$ may also indicate the possibility of a bifurcation leading to two different paths in the non-linear evolution, in which the left branch of figures 5(a) and 5(b) may lead to some kind of creasing instability like the one observed by Liang \& Cai (2015), or to some other new instability.
\begin{figure}
\centering
\begin{center}
\vspace*{-2 mm}
\includegraphics[width= 1.0 \textwidth, clip]{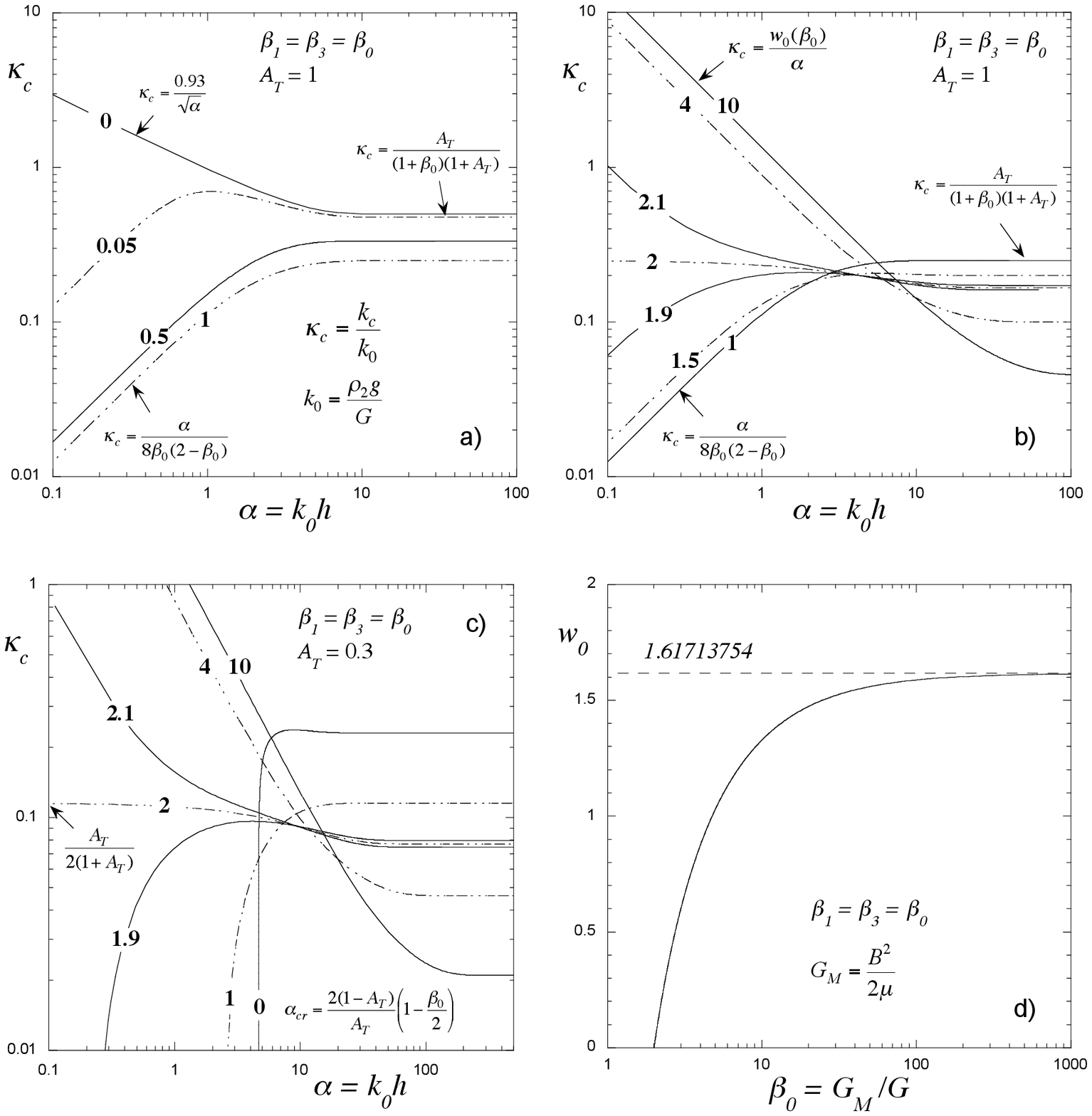}
\end{center}
\vspace*{-55 mm}
\caption{Marginal stability curves for the case with equal magnetic field intensities atop and beneath the slab, $\beta_1=\beta_3\equiv \beta_0$ [$\beta_n=B_{n}^{2}/(2\mu_nG)$], for two Atwood numbers ($A_T=1$ and $A_T=0.3$), and for several values of $\beta_0$ (indicated by the labels on the curves). (a) and (b) Dimensionless cut-off wavenumber $\kappa_c=w/\alpha$ as a function of $\alpha$ for $A_T=1$ and for $\beta_0 \leq 1$, and $\beta_0 \geq1$, respectively. (c) Dimensionless cut-off wavenumber $\kappa_c=w/\alpha$ as a function of $\alpha$ for $A_T=0.3$. (d) The function $w_0=w_0(\beta_0)$ given by (\ref{98}) for which $\alpha(w_0)=0$.}
\label{Figure-1}
\end{figure}

Besides, it is interesting to note that in this case the system stability increases with the intensity of the magnetic field $B_3$, such as shown by (\ref{87}) and figures 5(c) and 5(d). Such a behaviour is just the opposite to the one discussed in 4.2.1 for $\beta_3=0$ and $\beta_1\geq0$. However, it can be qualitatively explained in the same terms as before. In fact, now the presence of the magnetic field reduces the deformation $\xi_b$ of the upper surface of the slab, and such a reduction leads to an increase of the total strain $(\xi_a-\xi_b)/h$ for the thinner slabs. Therefore, the stabilising effect of elasticity is now reinforced by the magnetic field. In the case of a thicker the thicker slab, on the other hand, the effect of the magnetic field atop the slab is not felt and the system behaves like a purely elastic semi-infinite medium laying atop an ideal fluid.

\subsubsection{$\beta_1 = \beta_3 \equiv \beta_0$}

In this case, (\ref{74}) and (\ref{75})  reduce to the following expressions:
\begin{equation}
H_1(w)= \frac{1-A_T}{1+A_T} \left\{ \frac{w\left[2w(1-\beta_0) +\sinh 2w\right]}{(\sinh w)^2} +2w\beta_0 \right\}
\ ,\label{91}
\end{equation}
\begin{equation}
H_2(w)= 4w^2\left[1-\frac{(1-\beta_0)^2w^2}{(\sinh w)^2}\right]+ 4 w^2 \beta_{0}^{2} +4w^2\beta_0 \frac{\left[2w(1-\beta_0) +\sinh 2w\right]}{(\sinh w)^2}
\ .\label{92}
\end{equation}

For $w\gg1$ these equations yield, respectively, $H_1 \approx 2w(1+\beta_0)$ and $H_2 =[2w(1-\beta_0)]^2$, so we get the usual limit for thick slabs:
\begin{equation}
\kappa_c \approx \frac{A_T}{(1+A_T)(1+\beta_0)}
\ . \label{93}
\end{equation}

In the opposite limit $w\ll1$, we get:
\begin{equation}
H_1(w) \approx 2 \frac{1-A_T}{1+A_T} \left[2-\beta_0(1-w) \right]
\ ,\label{94}
\end{equation}
\begin{equation}
H_2(w)\approx 4 w^2 \left[1- (1-\beta_0)^2(1-\frac{w^2}{3}) +\beta_{0}^{2} \right] +8w\beta_0(2-\beta_0)
\ .\label{95}
\end{equation}
These limit show that (\ref{73}) describes several different behaviours depending on the values of $\beta_0$ and $A_T$ which are discussed below:

(i) For $A_T=1$ and $\beta_0=0$ we retrieve, as expected, the pure elastic case with a cut-off given by (\ref{81}) (Bakhrakh et al.1997, Plohr \& Sharp 1998, Piriz, Piriz \& Tahir 2017a, 2017b) [see figure 6(a), and (\ref{81})]

(ii) For $A_T=1$, and $0<\beta_0<2$, it turns out $H_1(w)=0$, and $H_2 \approx 8w\beta_0(2-\beta_0)$, and we get:
\begin{equation}
\kappa_c = \frac{\alpha}{8\beta_0(2-\beta_0)}
\ .\label{96}
\end{equation}
This shows that $\kappa_c$ decreases as $\beta_0$ increases provided that $\beta_0\leq1$ [figure 6(a)]. In this case, the behaviour is qualitatively similar to that one observed for the case with $\beta_1=0$ [figure 5(a)], indicating that for the smallest values of the magnetic fields the system response is dominated by the field atop the slab (for $\alpha \ll 1$).

Instead, for $\beta_0>1$ [figure 6(b)] the behaviour is inverted and $\kappa_c$ increases as the magnetic field increases ($\beta_0 <2$). This case, instead, is similar to the one for $\beta_3=0$ shown in figure 2(d), indicating that as $\beta_0$ increases beyond $\beta_0>1$ the stability system becomes dominated by the magnetic field beneath the slab.

(iii) For $\beta_0=2$ and arbitrary $A_T$, we have $H_1(w)\approx 2w\beta_0 = 4w$, and $H_2(w)\approx 4w^2\beta_{0}^{2}=(4w)^2$ ($w \ll1$), and the cut-off wavenumber reads [figures 6(b) and 6(c)]:
\begin{equation}
\kappa_c = \frac{A_T}{2(1+A_T)}
\ .\label{97}
\end{equation}
This value, for $A_T=1$, is larger than the one given by (\ref{96}) for $\beta_0<2$, indicating that $\kappa_c$ continues to increase with $\beta_0$.

(iv) For $\beta_0 >2$ and arbitrary $A_T$, we can see that exists a value $w_0$ of $w$ such that $\alpha (w_0)=0$. From (\ref{73}) we can see that it means that $H_2(w)=0$. This condition is only satisfied for some particular values of $\beta_0$, so that the function $w_0(\beta_0)$ is given by the implicit function obtained by solving the equation $H_2(w_0)=0$:
\begin{equation}
\beta_0(w_0)= \frac{(w_0+\sinh w_0)(1+\cosh w_0)}{w_0(1+cosh w_0)-(\sinh w_0)^2}
\ .\label{98}
\end{equation}
Actually $\beta_0(w_0)$ is more easily obtained by putting $C_0=A_0$ ($C_1=C_2\equiv C_0$ and $A_1=A_2\equiv A_0$) in (\ref{63}) and (\ref{64}). The function $w_0(\beta_0)$ has been represented in figure 6(d), and it shows that $w_0(\beta_0)$ has an asymptotic value $w_{0\infty}\approx1.616$ for $\beta_0 \rightarrow \infty$. Then, for $w_0=w_0(\beta_0)$ it is $\alpha \ll1$ and the cut-off wavenumber reads:
\begin{equation}
\kappa_c=\frac{w_0(\beta_0)}{\alpha}
\ ,\label{99}
\end{equation}
which is again larger than the value given by (\ref{97}) for $\beta_0=2$, and $\kappa_c$ continues to increase with $\beta_0$ up to achieve the asymptotic value $\kappa_{c \infty}\approx 1.616/\alpha$.

(v) For $0\leq \beta_0 \leq2$, and $A_T\neq1$, we have, in the limit $w\ll1$, $H_1(w)\approx 2(2-\beta_0)$ and $H_2(w)\approx 8\beta_0(1-\beta_0)w$. Thus, (\ref{73}) leads to the existence of a critical value $\alpha_{cr}$ for the slab instability, so that it is stable for $\alpha \leq \alpha_{cr}$:
\begin{equation}
\alpha_{cr}= \frac{2(1-A_T)}{A_T}\left(1-\frac{\beta_0}{2} \right)
\ .\label{100}
\end{equation}
Similarly to the case $\beta_3=0$ discussed in 4.2.1, there is an instability threshold below which the slab is stable and such a threshold progressively reduces as $\beta_0$ approaches to to the value $\beta_0=2$, for which it is completely removed. 

Therefore, we see that for $A_T\neq1$ the cut-off wavenumber monotonically increases with the intensity of the magnetic fields, for the thinner slabs, in a similar manner as we have seen for the case $\beta_3=0$. This indicates that the main conclusions obtained in such a case are still valid when $\beta_1 \sim \beta_3$.

\subsection{instability growth rate}

We can obtain the instability growth rate $\gamma$ from (\ref{63}) to $(\ref{69})$ as a function of the perturbation wavenumber $k$, in terms of the thickness $h$, the density $\rho_2$, and the shear modulus $G$ of the elastic slab, the density $\rho_1$ of the light medium beneath the slab, and the magnetic fields $B_1$ and $B_3$. For this, it is more convenient to use dimensionless magnitudes defined in (\ref{49a}).

Thus, after some tedious but straightforward algebra, we get the following implicit equation for $\sigma(\kappa)$, with the parameters $A_T$, $\alpha$, $\beta_1$, and $\beta_2$ already defined in previous sections:
\begin{eqnarray}
(1-\beta_1)(1-\beta_2) \Bigl\{(2\kappa^2+\sigma^2)^4 +16 \kappa^6(\kappa^2+\sigma^2) \Bigr.  \ \ \ \ \ \ \ \ \ \ \ \ \ \ \ \ \ \ \ \ \ \ \ \ \ \ \ \ \ \ \ \ \ \ \ \ \ \ \ \  \ \ \ \ \ \ \ \ \ \ 
\nonumber\\
  -
\left. 8\kappa^3\sqrt{\kappa^2+\sigma^2}(2\kappa^2+\sigma^2)^2 \left[\coth \alpha \kappa \coth \alpha \sqrt{\kappa^2+\sigma^2} -\mathrm{csch}\alpha \kappa \ \mathrm{csch}\alpha\sqrt{\kappa^2+\sigma^2} \right] \right\}  
   \nonumber\\
+
\beta_1 \beta_2 \sigma^4 \left\{ \left[(2\kappa^2+\sigma^2) \coth \alpha \kappa +2\kappa^2 \right]^2 - \left[(2\kappa^2+\sigma^2) \  \mathrm{csch} \alpha \kappa \right]^2 \right\}   
   \nonumber\\
+
[\beta_1(1-\beta_3)+\beta_3(1-\beta_1)] \sigma^2 \left\{ \mathcal{C} \left[ (2\kappa^2+\sigma^2)\coth \alpha \kappa +2\kappa^2 \right] - \mathcal{A} \left[(2\kappa^2+\sigma^2) \  \mathrm{csch} \alpha \kappa \right] \right\}
   \nonumber\\
   =
\kappa^2 \sigma^4-\frac{1-A_T}{1+A_T}\sigma^2(\kappa+\sigma^2) \left\{\kappa \sigma^2+(1-\beta_3)\mathcal{C}+\beta_3\sigma^2 \left[ (2\kappa^2+\sigma^2)\coth \alpha \kappa +2\kappa^2 \right] \right\}  
\nonumber\\
-
\kappa \sigma^2(\beta_3-\beta_1)\left\{ \mathcal{C} -\sigma^2 \left[ (2\kappa^2+\sigma^2)\coth \alpha \kappa +2\kappa^2 \right] \right\}       
, \label{102}  \ \ \ \ \ \ \  \ \ \ \ \ \ \ \ \ \ 
\end{eqnarray}
where
\begin{equation}
\mathcal{A}=(2\kappa^2+\sigma^2)^2 \ \mathrm{csch} \ \alpha \kappa- 4\kappa^3 \sqrt{\kappa^2+\sigma^2} \ \mathrm{csch} \ \alpha \sqrt{\kappa^2+\sigma^2}
\ . \label{103}
\end{equation}
\begin{equation}
\mathcal{C}=(2\kappa^2+\sigma^2)^2 \coth \alpha \kappa- 4\kappa^3 \sqrt{\kappa^2+\sigma^2} \coth \alpha \sqrt{\kappa^2+\sigma^2}
\ , \label{104}
\end{equation}

The expression (\ref{102}) is a bi-quartic transcendental equation that can be shown to have a unique real and positive root for any values of the arguments when the slab is unstable. Besides, it can be shown that $\sigma^2$ is always a real number, so that solutions with oscillating perturbations growth (over-stability) do not exist (see Appendix A). On the other hand, it means that for the unstable cases ($k<k_c$) there is a growing exponential mode of the form $e^{+|\gamma| t}$, and a decaying mode of the form $e^{-|\gamma| t}$ which together determine the evolution of the initial transient phase of growth for given initial conditions.Therefore, the dominant mode $|\sigma|$ given by (\ref{102}) is sufficient to characterise all the possible solutions.

For obtaining \ref{102} from \ref{69} we have done the algebra by hand and have verified it using the \textsc{Mathematica} software for symbolic calculations (Wolfram Research, Inc., 2015). The same procedure has been used in the previous long algebraic manipulations.

We have represented $\sigma(\kappa)$ for two different Atwood numbers ($A_T=1$ and $A_T=0.3$), and  for the three cases considered in the \S \ 4.2.1-3.
\begin{figure}
\centering
\begin{center}
\vspace*{-2 mm}
\includegraphics[width= 1.0 \textwidth, clip]{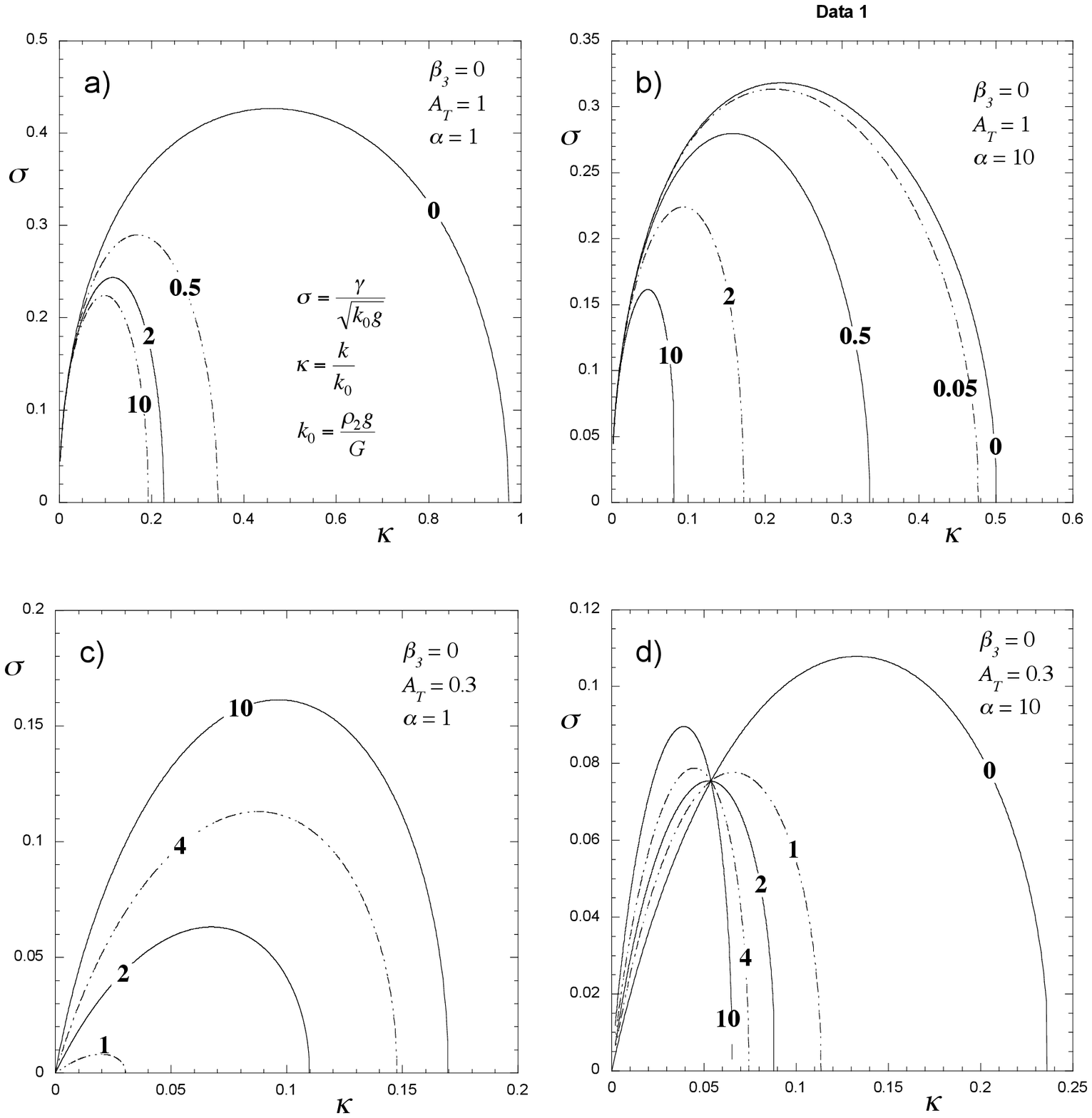}
\end{center}
\vspace*{-55 mm}
\caption{Dimensionless growth rate $\sigma = \gamma/\sqrt{k_0g}$ as a function of the dimensionless wavenumber $\kappa = k/k_0$, for the case with no magnetic field atop the slab ($\beta_3=0$), for two Atwood numbers $A_T$, for different values of the dimensionless magnetic pressure $\beta_1$ (indicated by the labels on the curves), and of the dimensionless slab thickness $\alpha$. (a) $A_T=1$ and $\alpha=1$. (b) $A_T=1$ and $\alpha=10$. (c) $A_T=0.3$ and $\alpha=1$. (d) $A_T=0.3$ and $\alpha=10$.}
\label{Figure-1}
\end{figure}

\begin{figure}
\centering
\begin{center}
\vspace*{-2 mm}
\includegraphics[width= 1.0 \textwidth, clip]{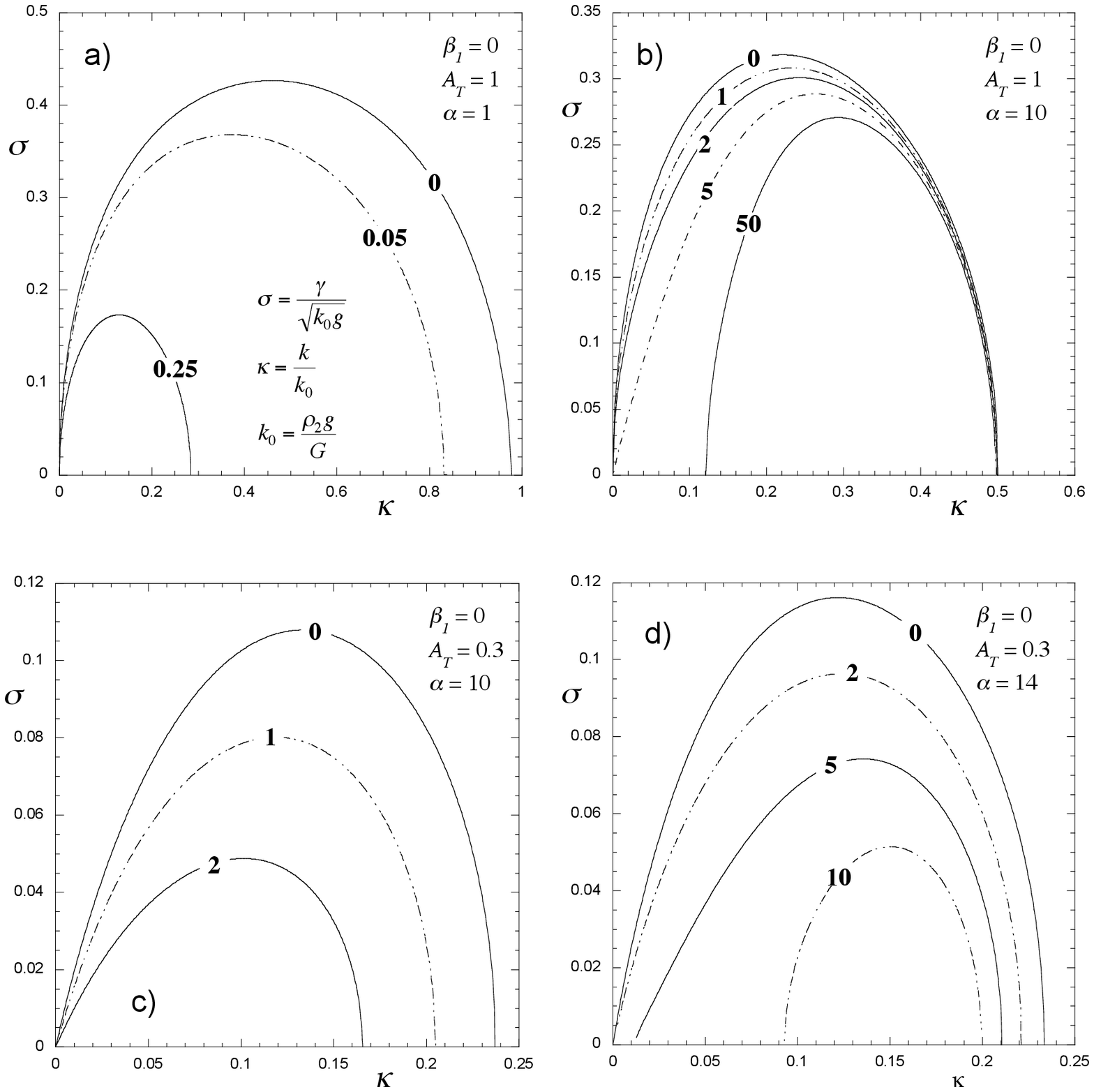}
\end{center}
\vspace*{-55 mm}
\caption{Dimensionless growth rate $\sigma = \gamma/\sqrt{k_0g}$ as a function of the dimensionless wavenumber $\kappa = k/k_0$, for the case with no magnetic field beneath the slab ($\beta_1=0$), for two Atwood numbers $A_T$, for different values of the dimensionless magnetic pressure $\beta_3$ (indicated by the labels on the curves), and of the dimensionless slab thickness $\alpha$. (a) $A_T=1$ and $\alpha=1$. (b) $A_T=1$ and $\alpha=10$. (c) $A_T=0.3$ and $\alpha=10$. (d) $A_T=0.3$ and $\alpha=14$.}
\label{Figure-1}
\end{figure}

\begin{figure}
\centering
\begin{center}
\vspace*{-2 mm}
\includegraphics[width= 1.0 \textwidth, clip]{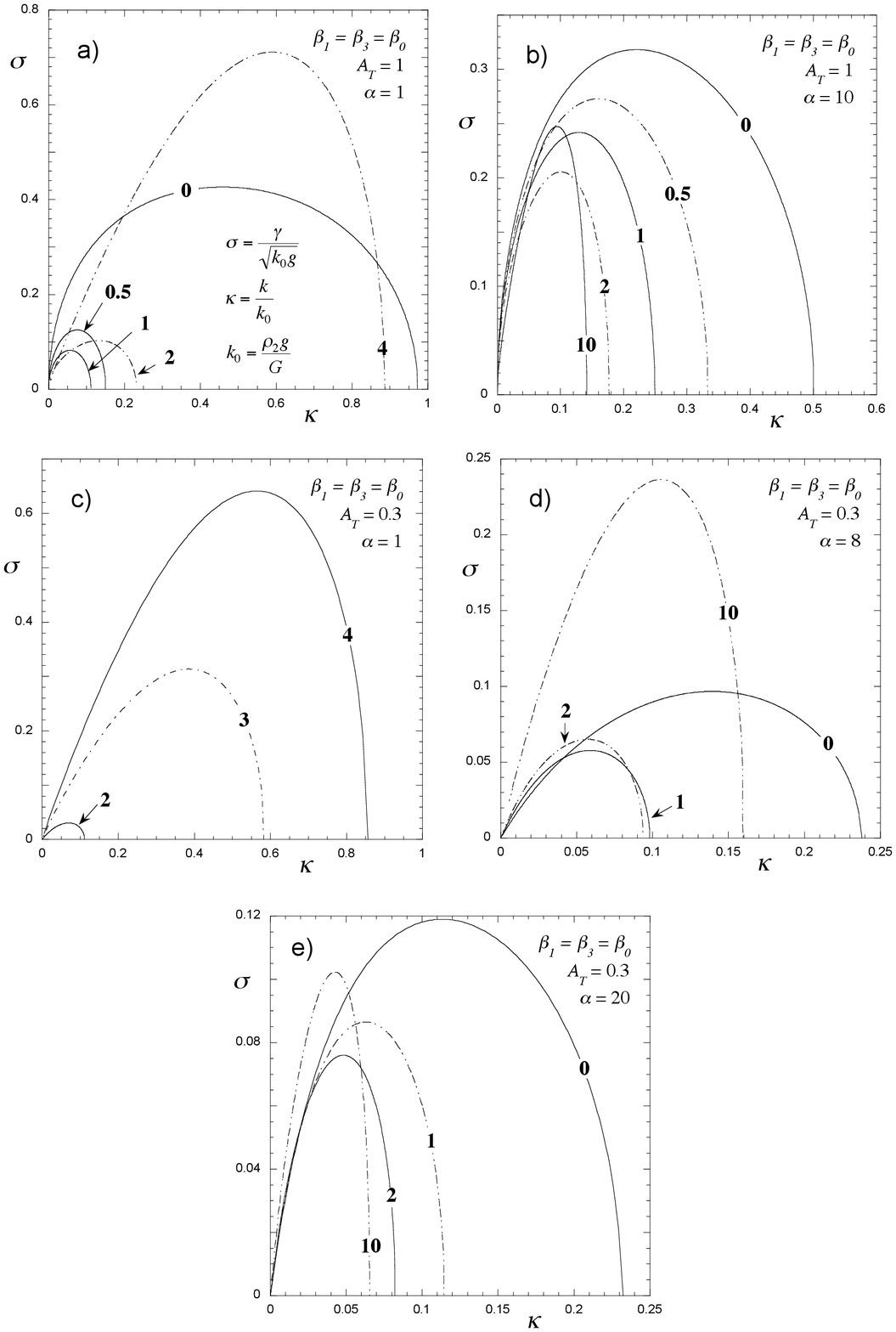}
\end{center}
\vspace*{-5 mm}
\caption{Dimensionless growth rate $\sigma = \gamma/\sqrt{k_0g}$ as a function of the dimensionless wavenumber $\kappa = k/k_0$, for the case with equal magnetic field intensities atop and beneath the slab, $\beta_1=\beta_3\equiv \beta_0$, for two Atwood numbers $A_T$, and for different values of the dimensionless magnetic pressures $\beta_0$ (indicated by the labels on the curves) and of the dimensionless slab thickness $\alpha$. (a)  $A_T=1$ and $\alpha=1$. (b) $A_T=1$ and $\alpha=10$. (c) $A_T=0.3$ and $\alpha=1$. (d) $A_T=0.3$ and $\alpha=8$. e) $A_T=0.3$ and $\alpha=20$.}
\label{Figure-1}
\end{figure}

\subsubsection{$\beta_1 \geq 0$, $\beta_3=0$}

Figures 7(a) and 7(b) show the instability growth rate for $A_T=1$ and for $\alpha=1$ and $\alpha=10$, respectively, and different values of $\beta_1$ indicated by the labels on the curves. These two values of $\alpha$ are representative of the two situations for thin and relatively thick slabs. By following the same tendency as the cut-off wavenumber, the maximum growth rate always decreases as the value of $\beta_1$ increases. But the growth rate reduction is less sensitive to the increase of $\beta_1$ for $\alpha=1$, especially for the largest values. We note that for $\alpha=10$ the asymptotic regime corresponding to a semi-infinite elastic medium has not yet been reached [see figure 4(c)], which is why the growth rate is somewhat higher than the one that would be obtained from (\ref{47}).

For $A_T=0.3$, on the other hand, the behaviour is quite different for the thin and the thick slabs, such as we have already seen in figure 4(d). In fact, figure 7(c) shows that for $\alpha=1$ the growth rate monotonically decreases with $\beta_1$, and it becomes $\gamma=0$ for $\beta_1 \leq 2(1-A_T)/(1+A_T)$. For $\alpha=10$ the growth rate does not decreases monotonically when $\beta_1$ increases as the cut-off wavenumber does [figure 7(d)], but it considerably decreases for $0\leq \beta_1\leq 1$ and then it remains more or less the same with some increase for the largest values of $\beta_1$. It is actually difficult to provide a clear physical explanation for such a behaviour since it depends on the interplay of the effects due to the presence of the lighter fluid with those of elasticity, magnetic field, and slab thickness that, as we have already seen, may compete among them.

\subsubsection{$\beta_1 = 0$, $\beta_3 \geq 0$}

In this case, the growth rate follows a monotonic behaviour for any value of $A_T$ and $\alpha$, in which $\sigma$ decreases as $\beta_3$ increases (figure 8). However, for $\beta_3>2$ ($\alpha >4$), a second cut-off appears from the side of the shortest wavenumbers that is clearly seen in figures 8(b) and 8(d) [see also figures 5(a) and 5(b)]. When it occurs, the classical growth rate for the RT instability ($\gamma \sim \sqrt{kg}$) usually expected for the small values of the wavenumber $k$, is not any longer retrieved. 

The existence of this short wavenumber cut-off is related to the behaviour already observed in figures 5(a) and 5(b), and,as it was discussed in \S \ 4.2.2, it is somewhat similar to what it was observed by Mora et al. (2014) and Ricobelli \& Ciarletta (2017) for an elastic slab in contact with a rigid wall. But, as we have already seen, although the effect of the magnetic field above  the slab is to reduce the vertical deformation of the slab top surface, it does not affect the tangential velocity. Therefore, rigid wall conditions are never retrieved even for $\beta_3 \rightarrow \infty$. 

On the other hand, the reduction of the deformation $\xi_b$ of top surface leads to an enhancement of the total relative deformation of the slab $(\xi_a-\xi_b)/h$ thereby reinforcing the stabilising elasticity effectivity. However, such an effect cannot be felt for the shortest perturbation wavelengths ($kh\gtrsim1$), which cannot "see" the slab top surface and whereby the total relative deformation is instead $k \xi_a$. AS a result, the stabilising effect of the magnetic field occupying the region $y\leq -h$ is only felt for the perturbation wavenumbers such that $kh \lesssim 1$, what leads to a short wavenumber cut-off.

\subsubsection{$\beta_1 = \beta_3 \equiv \beta_0$}

As is seen in \S \ 4.2.3, this case presents some mixed characteristics of the two previous cases discussed above. In figures 9(a) and 8(b) we show the growth rate for $A_T=1$ and $\alpha=1$ and $\alpha=10$, respectively. As in the previous cases, we also considerfor several values of $\beta_0$ indicated by the labels on the curves. The maximum growth rate follows the same tendency as the cut-off, namely it decreases as $\beta_0$ increases, provided that $\beta_0 <1$, while inverting this behaviour in the opposite case [see (\ref{96})]. For $\alpha=10$, $\sigma$ again follows the behaviour of $\kappa_c$ [figures 6(a) and 6(b)] for relatively weak magnetic fields and it decreases as $\beta_0$ increases. But for the largest values of $\beta_0$ the growth rate starts to increase. It is not completely clear what is the physical reason for this loss of stability for the most intense magnetic fields occurring in the regime of  intermediate wavenumbers, and it is difficult to figure out the details of the interplay among the different effects. But it seems that for the largest values of $\beta_0$ the magnetic field beneath the slab becomes more relevant for determining the slab stability. 

For $A_T=0.3$, we have already seen in \S \ 4.2.3 that the cut-off wavenumber always increases with $\beta_0$ provided that the slab is sufficiently thin. And, according to (\ref{100}), it becomes zero for a given value of $\alpha$ when $\beta_0 \leq [2-\alpha A_T/(1-A_T)]$. Figure 9(c) shows that for $\alpha=1$,  the growth rate follows the same tendency than the cut-off. But in figure 9(d) we see that the behaviour is the contrary for very large values of $\alpha$, approaching to the case of two semi-infinite media. 

However, for the intermediate values of $\alpha$ the interplay among the different mechanisms makes the growth rate to exhibit a variety of behaviours surely depending of the relative dominance of each one of them [figures 9(d) and 9(e)]. As we have already discussed, we can only provide physical interpretations for the extreme cases for which some specific mechanisms is seen to be dominant over the others.

\section{Concluding remarks}

We have presented a linear theory for the two-dimensional MRT instability in a system that is composed of an elastic layer that lies above a lighter ideal fluid. Moreover, a uniform magnetic filed is present above and below this system, as shown in figure 1. Consideration of a finite thickness of the elastic layer in this work, lead to the discovery of interesting new aspects of this problem, which have not been detected in the previous studies that involved semi-infinite media, and/or do not consider Hookean constitutive properties of the slab.

The magnetic field which exists in the region occupied by the lighter fluid, could be expected on the basis of the results for semi-infinite media, to provide a positive and supporting contribution to the stabilising effect produced by the elasticity (addition of the two effects). However, contrary to this expectation, the magnetic field opposes the elasticity stabilization effect when the layer is sufficiently thin. This is because for relatively thin layers, the total strain that controls the elasticity effects, is determined by the deformation of both slab interfaces, whereas, the magnetic field affects mainly the face on which it is acting (see discussion in \S \ 4.2.1). As a consequence, the influence of the magnetic field acting below the elastic layer becomes detrimental to the stabilising effects due to the elasticity which reduces the system stability.  Furthermore, the instability threshold imposed by the layer elasticity is progressively reduced and even vanishes for a sufficiently large magnetic pressure ($\sim G$).

This situation is very common in many high energy density physics experiments and inertial fusion schemes in which a finite thickness slab is accelerated, or a cylindrical shell target is imploded by a magnetic pressure. The slab is maintained in a solid state with the aim to enhance the hydrodynamic stability during acceleration. The new results that we report may be an important issue in such experiments. 
 
In Nature, the present problem is also very relevant to the triggering of crust-quakes in the strongly magnetised neutron stars known as megnetars. The mechanism proposed by Blaes et al. (1990, 1992) requires a minimum inversion density in the crust in order to overtake the threshold imposed by the elasticity. It has been shown that such an inversion density can occur as a consequence of the pycnonuclear and electron capture reactions forced by the neutron-star matter-accretion from the interstellar medium. Nevertheless, it seems quite improbable that it may have enough magnitude to exceed the instability threshold established by the crust elasticity. However, the presence of the strong magnetic fields generating magnetic pressures of the order of the shear modulus $G$ can eliminate such a threshold and make the crust unstable for any arbitrary small density inversion. Therefore, MRT instability can be an effective process to trigger crust-quakes in magnetars when magnetic pressures of the order of the shear modulus, G, of the crust are developed.

In a similar manner, the magnetic field on top of the elastic layer acts in support of the stabilising effects generated by elasticity, but its action on the system is limited to the relatively long perturbation wavelengths ($kh \gtrsim 1$). As a consequence, a short wavenumber cut-off may exist, below which the system remains stable. This case possess some similarities with the RT instability in elastic layers in contact with rigid walls. This is because such a magnetic field restricts the velocity perturbation normal to the interface, while  the tangential velocity is not affected and therefore the rigid wall boundary conditions are never reproduced. Nevertheless, we can speculate on the possibility that in the non-linear regime the system may evolve towards two different instabilities (creasing and wrinkle instabilities) corresponding to each branch of short and long wavenumbers, as reported by Liang \& Cai (2015) for the case of elastic soft materials.

On the other hand, when magnetic fields on both sides of the elastic slab are comparable, the resulting picture is a combination of the previous two extreme cases. In general, for the shortest perturbation wavelengths, the effects of the magnetic field beneath the slab becomes dominant, and the effects of the field atop it are felt for the thinner slabs or the longest perturbation wavelengths.

Finally, it may be worth to remark that the present linear theory assumes that the deviation from flatness of the slab is always sufficiently small, so that it must be $k\xi_{a,b} \ll 1$.

\section*{Acknowledgements}
This work has been partially supported by Ministerio de Econom\'{i}a  y Competitividad (Grant No. ENE2016-75703-R) and Junta de Comunidades de Castilla-La Mancha of Spain (Grant No. SBPLY/17/180501/000264), and by the BMBF of Germany. 

\appendix
\section{Proof that $\sigma^2$ is a real number}
\label{appA}

We start with (\ref{18}$b$) and its complex conjugate:
\refstepcounter{equation}
$$
\gamma^2 \psi_2=\frac{G}{\rho_2} \nabla^2 \psi_2, \quad
(\gamma^2)^* \psi_2=\frac{G}{\rho_2} \nabla^2 \psi_{2}^{*} \ ,
  \eqno{(\theequation{\mathit{a},\mathit{b}})} \label{a1}
$$
where $(\gamma^2)^*$ and $\psi_{2}^{*}$ are the complex conjugate of $\gamma^2$ and $\psi_{2}$, respectively. By multiplying the first one by  $\psi_{2}^{*}$ and the second one by  $\psi_{2}$, and subtracting, we get:
\begin{equation}
\left[\gamma^2-(\gamma^2)^* \right] |\psi_2|^2= \frac{G}{\rho_2}\left[\psi_{2}^{*}  \nabla^2 \psi_{2}- \psi_{2}  \nabla^2 \psi_{2}^{*} \right] = \frac{G}{\rho_2} \boldsymbol{\nabla \cdot} \left( \psi_{2}^{*}  \boldsymbol{\nabla}  \psi_{2}- \psi_{2} \boldsymbol{\nabla} \psi_{2}^{*} \right)
\ .\label{a2}
\end{equation}
By integrating over the two-dimensional volume $V= h \ell$ ($\ell=2 \pi/k$ is the perturbation wavelength), and then using the Green's theorem to transform the volume integral into a surface integral over the surface $A(V)$ of such a volume, it yields:
\begin{equation}
\left[\gamma^2-(\gamma^2)^* \right] \int_V |\psi_2|^2 \ dV=  \frac{G}{\rho_2} \int_{A(V)}  \left( \psi_{2}^{*}  \boldsymbol{\nabla}  \psi_{2}- \psi_{2} \boldsymbol{\nabla} \psi_{2}^{*} \right)\boldsymbol{\cdot} d\boldsymbol{A} 
\ .\label{a3}
\end{equation}
By performing the surface integral by pieces over the surface $A(V)$, we have:
\begin{eqnarray}
 \left[\gamma^2-(\gamma^2)^* \right] \int_V |\psi_2|^2 \ dV=  \frac{G}{\rho_2} \int_{0}^{h} \left. \left( \psi_{2}^{*}  \frac{\partial \psi_{2}}{\partial x}- \psi_{2}  \frac{\partial \psi_{2}^{*}}{\partial x} \right)\right|_{x=0}^{x=\ell} dy  \ \ \ \ \ \  \ \ \ \ \ \  \ \ \ \ \ \  
  \nonumber\\
 + 
 \int_{0}^{\ell} \left.\left( \psi_{2}^{*}  \frac{\partial \psi_{2}}{\partial y}- \psi_{2}  \frac{\partial \psi_{2}^{*}}{\partial y} \right)\right|_{y=0}^{y=h} dx
. \label{a4}
\end{eqnarray}
Since from \ref{20} we have:
\begin{equation}
\psi_2 = e^{\gamma t}f(y) \cos kx \  \  ,  \  \  \  \  \psi_{2}^{*} = e^{\gamma^* t}f^*(y) \cos kx
\  , \label{a5}
\end{equation}
it is straightforward to see that the integrand of the first integral is identically zero, and that the second integral is proportional to $\int_{0}^{\ell} \cos kx \ dx =0$. Therefore, it turns out that $\gamma^2=(\gamma^2)^*$ and $\sigma^2$ is a real number. That is, there are no oscillating growth solutions, and \ref{102} gives all possible solutions for $\gamma(k)$.

\end{document}